\documentclass[a4paper,12pt]{article}
\usepackage{cite}
\usepackage[utf8x]{inputenc}
\usepackage{amsmath}
\usepackage{amsfonts}
\usepackage{amssymb}
\usepackage{ctable}
\usepackage{subfigure}
\usepackage{graphicx} 
\usepackage[hyperindex=true,breaklinks]{hyperref}
\usepackage[english]{babel}
\usepackage[hmargin=2.0cm,vmargin=3.0cm]{geometry}
\DeclareMathOperator{\sech}{sech}

\usepackage{amsfonts}

\begin{document} 

\title{Description of High-Energy pp Collisions Using Tsallis Thermodynamics: Transverse Momentum and Rapidity Distributions}
\author{L. Marques$^1$, J. Cleymans$^2$ and A. Deppman$^1$}
\date{\small$^1$ Instituto de F\'isica, Universidade de S\~ao Paulo - IFUSP, Rua do Mat\~ao, Travessa R 187, 
05508-900 S\~ao Paulo-SP, Brazil \\ $^2$ UCT-CERN Research Centre and Department of Physics, University of Cape Town, Rondebosch 7701, South Africa}

\maketitle

\begin{abstract}
A systematic analysis of transverse momentum and rapidity distributions measured in 
high-energy  proton - proton (pp) collisions for energies ranging from 53 GeV   to 7 TeV using Tsallis thermodynamics is presented. 
The  excellent description of all transverse momentum spectra obtained in earlier analyses is confirmed and extended.
All energies can be described by  a  single Tsallis temperature of 68 $\pm$ 5  MeV at all beam energies and particle types investigated (43 in total).
The value of the entropic index, q, shows a wider spread but is always close to q $\approx$ 1.146. These values are then used 
to describe the rapidity distributions using a superposition of two Tsallis fireballs along the rapidity axis. 

It is concluded that the hadronic system created in high-energy p - p collisions between 53 GeV and 7 TeV 
can be seen as obeying Tsallis thermodynamics.
\end{abstract}
\section{Introduction}
Recently a power-law function based on  the  Tsallis distribution has been used extensively in fits of the $p_T$
distributions measured in high-energy collisions~\cite{STAR, PHENIX,ATLAS,ALICE,CMS}.
The relationship with an 
approach based on Tsallis thermodynamics~\cite{Tsallis1988,Tsallis1999,biro} 
(and not simply the distribution associated with it) 
has been clarified  in several papers~\cite{Beck,Deppman,CleymansWorku,Azmi}, in particular, the exponent appearing  in the probability 
distribution used in~\cite{STAR, PHENIX,ATLAS,ALICE,CMS} is associated  with the entropic index, $q$ forming the basis of Tsallis thermodynamics. 
A further step  was made in~\cite{Deppman} to generalize Hagedorn's theory~\cite{Hagedorn65,Hagedorn83} to  non-extensive thermodynamics that 
predicts a limiting temperature ($T$) and a characteristic entropic index for hadronic 
systems~\cite{Deppman}. 

In this paper a systematic analysis of transverse momentum and rapidity distributions measured in 
high-energy  proton - proton (pp) collisions for energies ranging from 53 GeV   to 7 TeV using Tsallis thermodynamics is presented. 
The  excellent description of all transverse momentum spectra obtained in earlier analyses is confirmed and extended.
The present analysis of $p_T$-distributions confirms previous results~\cite{Sena1,Sena2,CleymansWorku,Azmi} 
giving constant temperature and constant entropic index when 
the non-extensive distribution for systematic analysis of experimental data. 
A recent analysis~\cite{wilk1,wilk2} obtained a higher temperature but it can be shown that this is due to not using a summation over all 
charged particles produced but simply fitting a single formula, doing so makes it consistent with the temperature $T$ used in this paper. 
In this paper  these values are used  to analyse rapidity distributions. 
In this case we use a model to describe the fireball rapidity distribution along the rapidity axis and show that a 
consistent pattern emerges which allow us to 
make extrapolations for the upcoming pp collision at the upgraded LHC.

In what follows we present a new analysis of $p_T$-distribution extending those already published, and 
provide more precise values for $T$ and $q$. Then we present a model based on the theory for describing rapidity distribution and 
perform a systematic analysis of experimental data on rapidity distribution of charged hadrons is performed. Finally the results 
of the analysis are used for predicting $p_T$ and rapidity distributions for the next LHC phase at 13 TeV.

\section{Transverse Momentum Distributions}
  
The power-law distribution based on Tsallis thermodynamics is given, in terms of transverse momentum and rapidity, by~\cite{CleymansWorku}
\begin{equation}
 \frac{d^2N}{p_T\,dp_T\,dy}=gV\frac{m_T \cosh y}{(2\pi)^2}\bigg[1+(q-1)\frac{m_T \cosh y-\mu}{T} \bigg]^{-\frac{q}{q-1}}
 \label{eq:doublediff}
\end{equation}
where $\mu$ is the chemical potential, $m_T=\sqrt{p_T^2+m_0^2}$, $V$ is the volume, $g$ is the degeneracy factor and $q$ is the 
entropic factor, which measures the non-additivity of the entropy. Boltzmann statistics is recovered with $q = 1$. This equation can be obtained also from a non-extensive version of the perfect gas partition function \cite{Megias}.

It has been shown that at central rapidity
 $y=0$ one can easily obtain the transverse momentum distribution 
in terms of the central rapidity density, $\frac{dN}{dy}\bigg|_{y=0}$, as~\cite{Poland}
\begin{equation}
 \frac{d^2N}{dp_Tdy}\bigg|_{y=0} = \frac{p_T\,m_T}{T} \frac{dN}{dy}\bigg|_{y=0} \frac{(2-q)(3-2q)}{(2-q)m_0^2 + 2m_0T + 2T^2}\bigg[1+(q-1)\frac{m_0}{T} \bigg]^{\frac{1}{q-1}} \bigg[1+(q-1)\frac{m_T}{T} \bigg]^{-\frac{q}{q-1}}
\label{fittingformula}
\end{equation}

The above equation will be used for a systematic analysis of $p_T$-distribution for pp collisions for a 
large set of experiments, as listed in Table \ref{tab:Tablept}. A subset of those experiments was already analysed in 
Ref.~\cite{Lucas}. Since all experiments report results for a narrow range of rapidity around the central region the 
approximation used is appropriate. The only free parameters to be adjusted to the experimental data are $q$ and $T$. 
According to the thermodynamical theory in Ref. \cite{Deppman}, if both the self-consistent principle from 
Hagedorn~\cite{Hagedorn65} and the Tsallis 
statistics~\cite{Tsallis1988, Tsallis1999,biro} can be applied in high-energy physics then both $T$ and $q$ must be independent of 
particle type. 

\begin{table}[!ht]\centering
\caption{ }
\label{tab:Tablept}
\scalebox{0.95}{
\begin{tabular}{c c c c c | c c c c c} \toprule
 Index    &   Particle    &$\sqrt{s}\,(TeV)$ & Exp.       &     Ref.         &Index     & Particle      & $\sqrt{s}\,(TeV)$  & Exp.      & Ref.              \\ \midrule
     1    & $\pi^+$       & 0.2              & PHENIX     & \cite{Artigo13}  &    22    & $P^-$         & 0.9                & CMS       & \cite{Artigo16}   \\
     2    & $\pi^-$       & 0.2              & PHENIX     & \cite{Artigo13}  &    23    & $\Lambda$     & 0.9                & CMS       & \cite{CMS}        \\
     3    & $K^+$         & 0.2              & PHENIX     & \cite{Artigo13}  &    24    & $\Xi^-$       & 0.9                & CMS       & \cite{CMS}        \\
     4    & $K^-$         & 0.2              & PHENIX     & \cite{Artigo13}  &    25    & $\pi^+$       & 2.76               & CMS       & \cite{Artigo16}   \\
     5    & $P^+$         & 0.2              & PHENIX     & \cite{Artigo13}  &    26    & $\pi^-$       & 2.76               & CMS       & \cite{Artigo16}   \\
     6    & $P^-$         & 0.2              & PHENIX     & \cite{Artigo13}  &    27    & $K^+$         & 2.76               & CMS       & \cite{Artigo16}   \\
     7    & $P^+$         & 0.2              & PHENIX     & \cite{Artigo13}  &    28    & $K^-$         & 2.76               & CMS       & \cite{Artigo16}   \\
     8    & $P^-$         & 0.2              & PHENIX     & \cite{Artigo13}  &    29    & $P^+$         & 2.76               & CMS       & \cite{Artigo16}   \\
     9    & $\pi^0$       & 0.9              & ALICE      & \cite{Artigo1}   &    30    & $P^-$         & 2.76               & CMS       & \cite{Artigo16}   \\
     10   & $\pi^+$       & 0.9              & ALICE      & \cite{ALICE}     &    31    & $\pi^0$       & 7.0                & ALICE     & \cite{Artigo1}    \\
     11   & $\pi^-$       & 0.9              & ALICE      & \cite{ALICE}     &    32    & $\pi^+$       & 7.0                & CMS       & \cite{Artigo16}   \\
     12   & $\pi^+$       & 0.9              & CMS        & \cite{Artigo16}  &    33    & $\pi^-$       & 7.0                & CMS       & \cite{Artigo16}   \\
     13   & $\pi^-$       & 0.9              & CMS        & \cite{Artigo16}  &    34    & $K^0_s$       & 7.0                & CMS       & \cite{CMS}        \\
     14   & $K^0_s$       & 0.9              & CMS        & \cite{CMS}       &    35    & $K^+$         & 7.0                & CMS       & \cite{Artigo16}   \\
     15   & $K^+$         & 0.9              & ALICE      & \cite{ALICE}     &    36    & $K^-$         & 7.0                & CMS       & \cite{Artigo16}   \\
     16   & $K^-$         & 0.9              & ALICE      & \cite{ALICE}     &    37    & $\eta$        & 7.0                & ALICE     & \cite{Artigo1}    \\
     17   & $K^+$         & 0.9              & CMS        & \cite{Artigo16}  &    38    & $K^*$         & 7.0                & ALICE     & \cite{Abelev2}    \\
     18   & $K^-$         & 0.9              & CMS        & \cite{Artigo16}  &    39    & $P^+$         & 7.0                & CMS       & \cite{Artigo16}   \\
     19   & $P^+$         & 0.9              & ALICE      & \cite{ALICE}     &    40    & $P^-$         & 7.0                & CMS       & \cite{Artigo16}   \\
     20   & $P^-$         & 0.9              & ALICE      & \cite{ALICE}     &    41    & $\phi$        & 7.0                & ALICE     & \cite{Abelev2}    \\
     21   & $P^+$         & 0.9              & CMS        & \cite{Artigo16}  &    42    & $\Lambda$     & 7.0                & CMS       & \cite{CMS}        \\ \bottomrule
\end{tabular} }
\end{table}

\begin{figure}[!ht]
   \centering
   \subfigure[]{
      \label{figFotoFiss:subfig2:a}
      \includegraphics[scale=0.38]{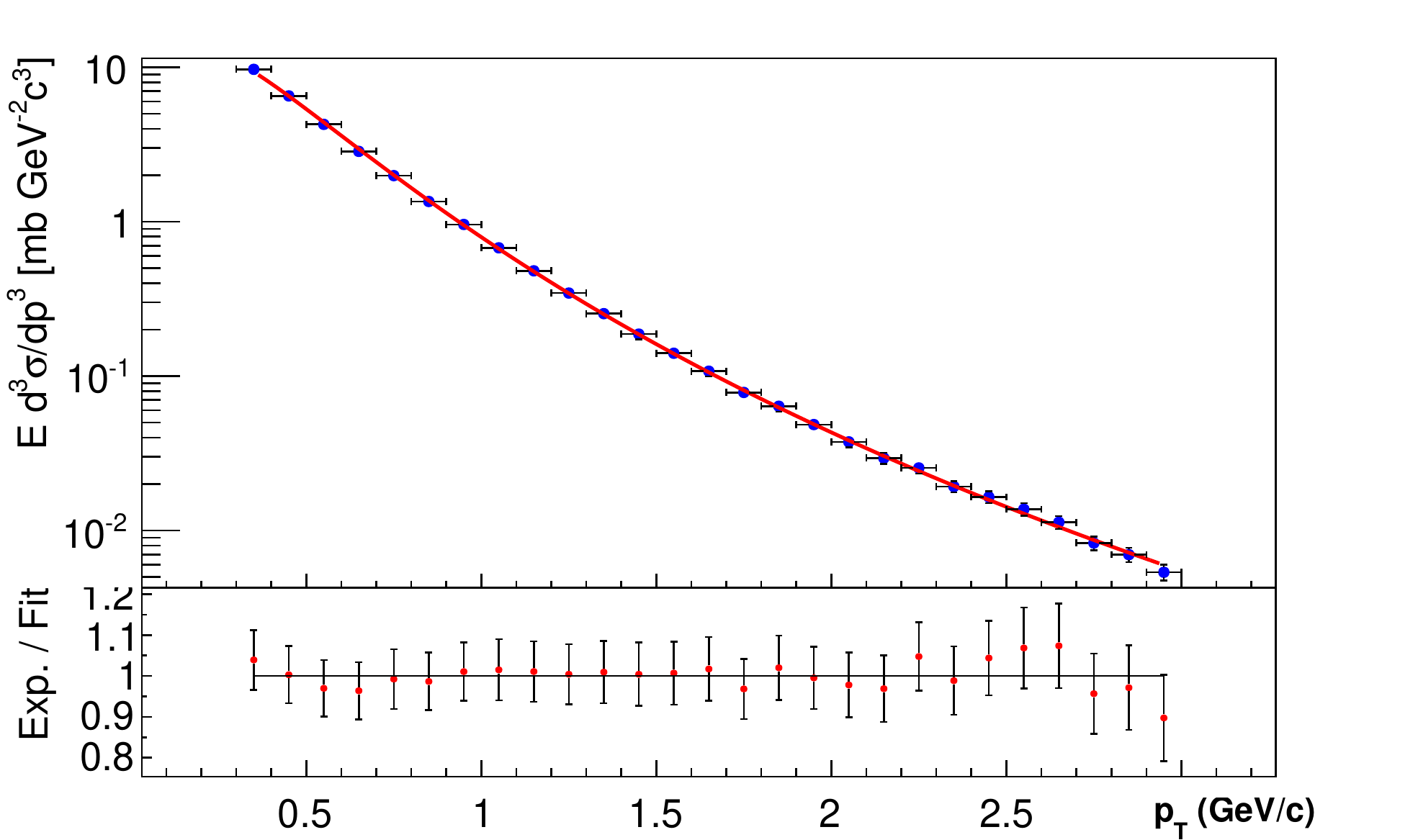}
   }
   \subfigure[]{
      \label{figFotoFiss:subfig2:b}
      \includegraphics[scale=0.38]{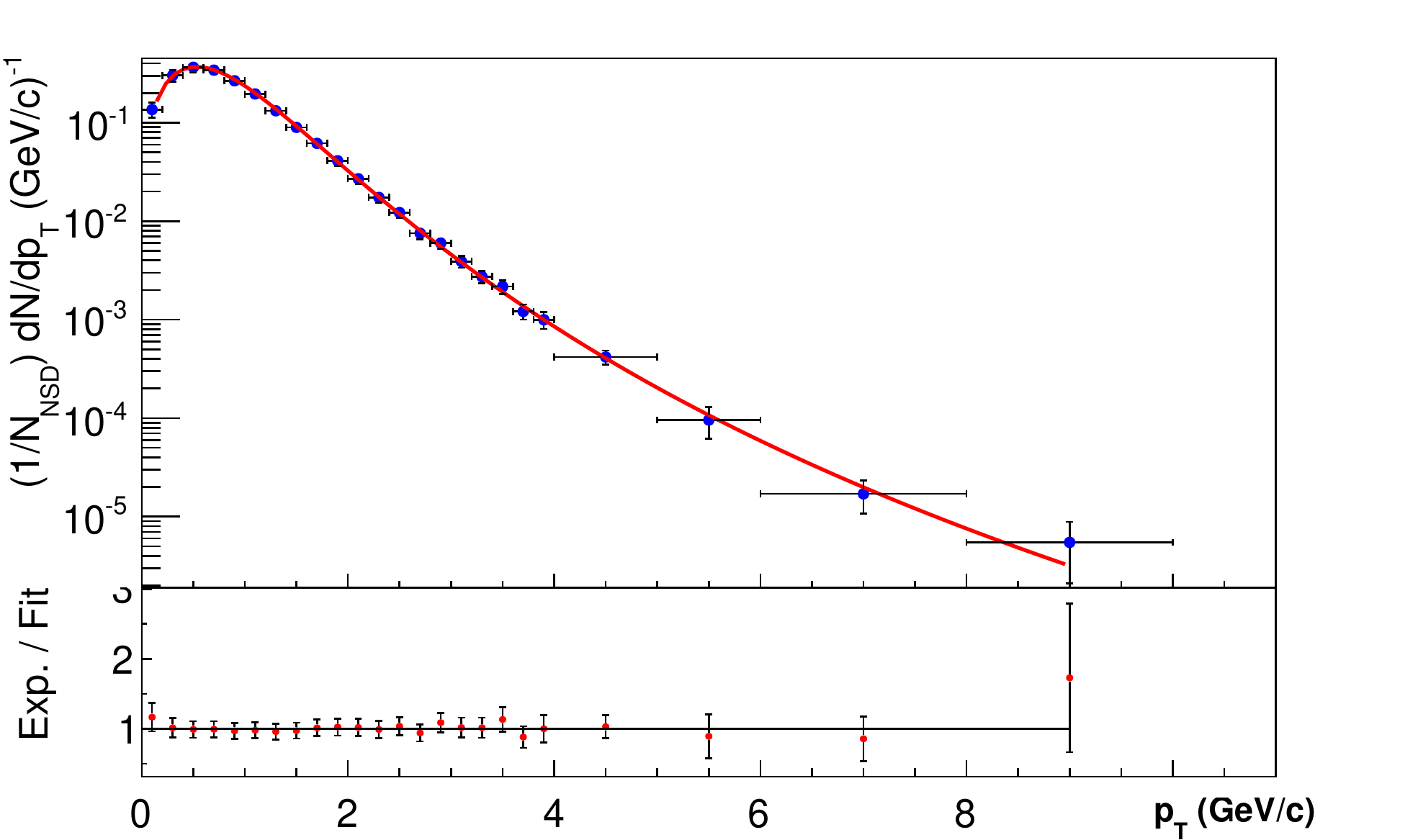}
   }
   \subfigure[]{
      \label{figFotoFiss:subfig2:c}
      \includegraphics[scale=0.38]{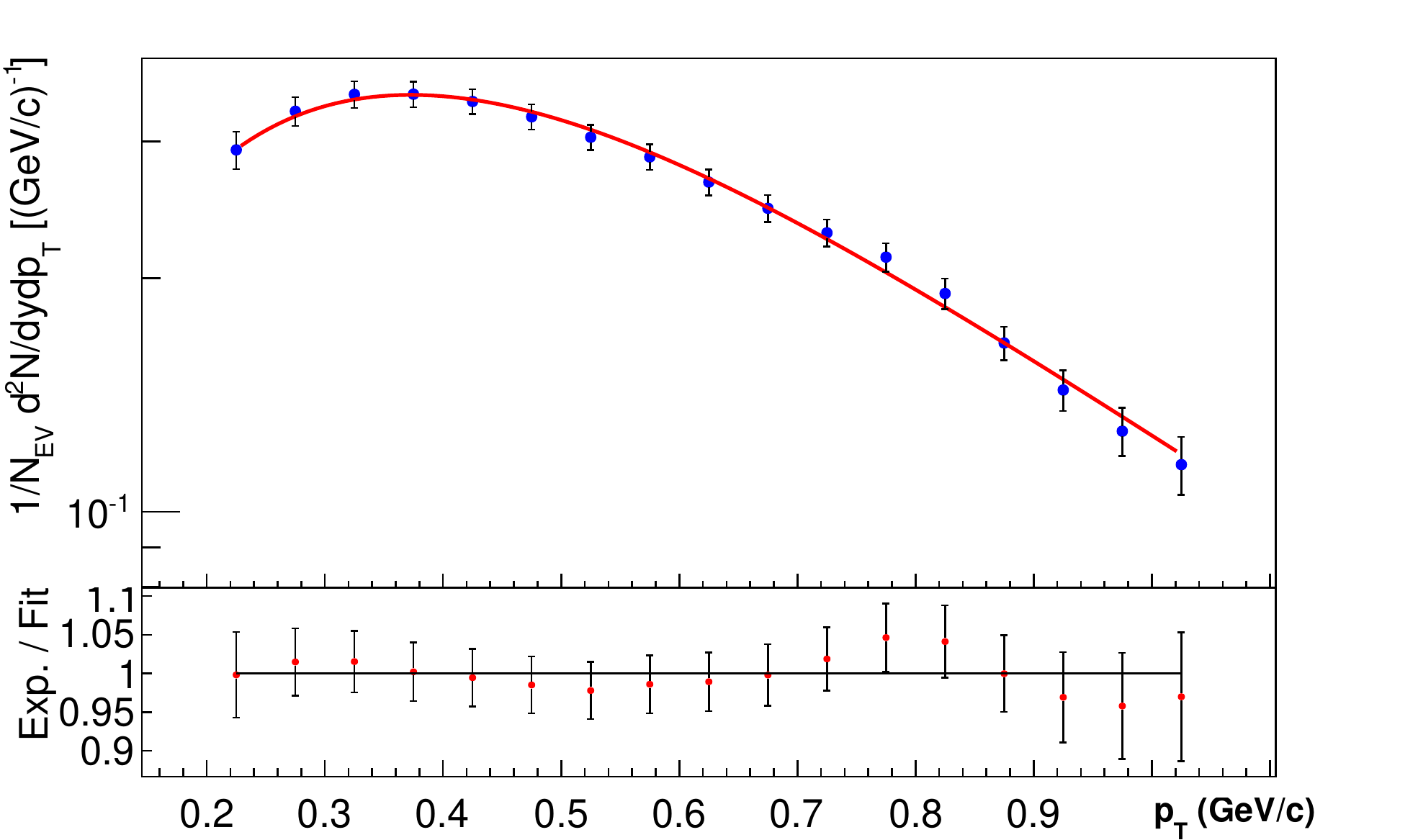}
   }
   \subfigure[]{
      \label{figFotoFiss:subfig2:d}
      \includegraphics[scale=0.38]{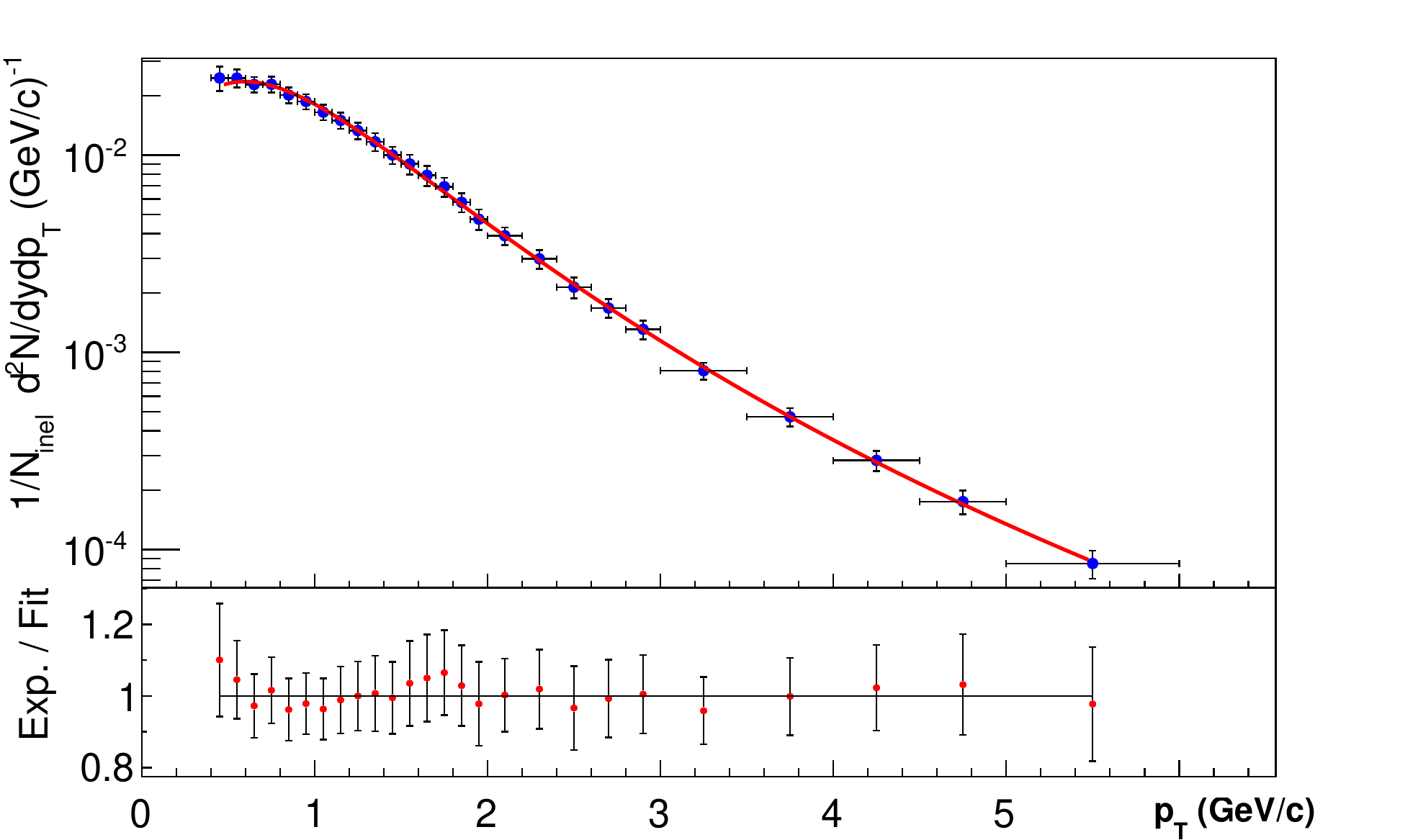}
   }
  \caption{Fits to the $p_T$-distribution given by Eq.~\ref{eq:fitformula} to experimental data. (a) $\pi^+$, $\sqrt{s} = 0.2$ TeV, \cite{Artigo13} (b) $\Lambda$, $\sqrt{s} = 0.9$ TeV, \cite{CMS} (c) $K^-$, $\sqrt{s} = 2.76$ TeV, \cite{Artigo16} and (d) $\phi$, $\sqrt{s} = 7.0$ TeV, \cite{Abelev2}.}
   \label{fig:ExAjustedN}
\end{figure}

\begin{figure}[!ht]
    \centering
    \subfigure[]{
	    \includegraphics[scale=0.60]{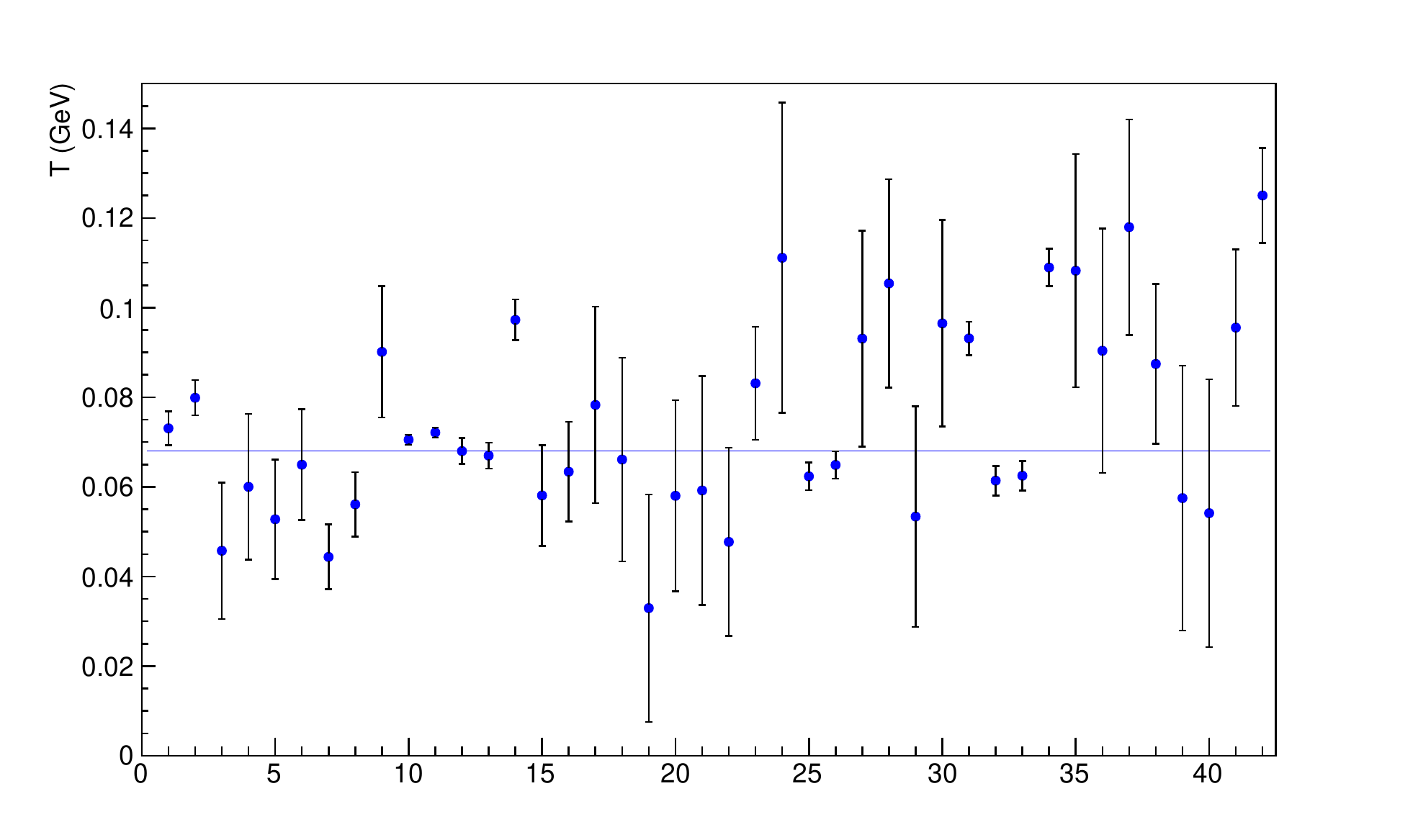}
    }
    \subfigure[]{
    \includegraphics[scale=0.60]{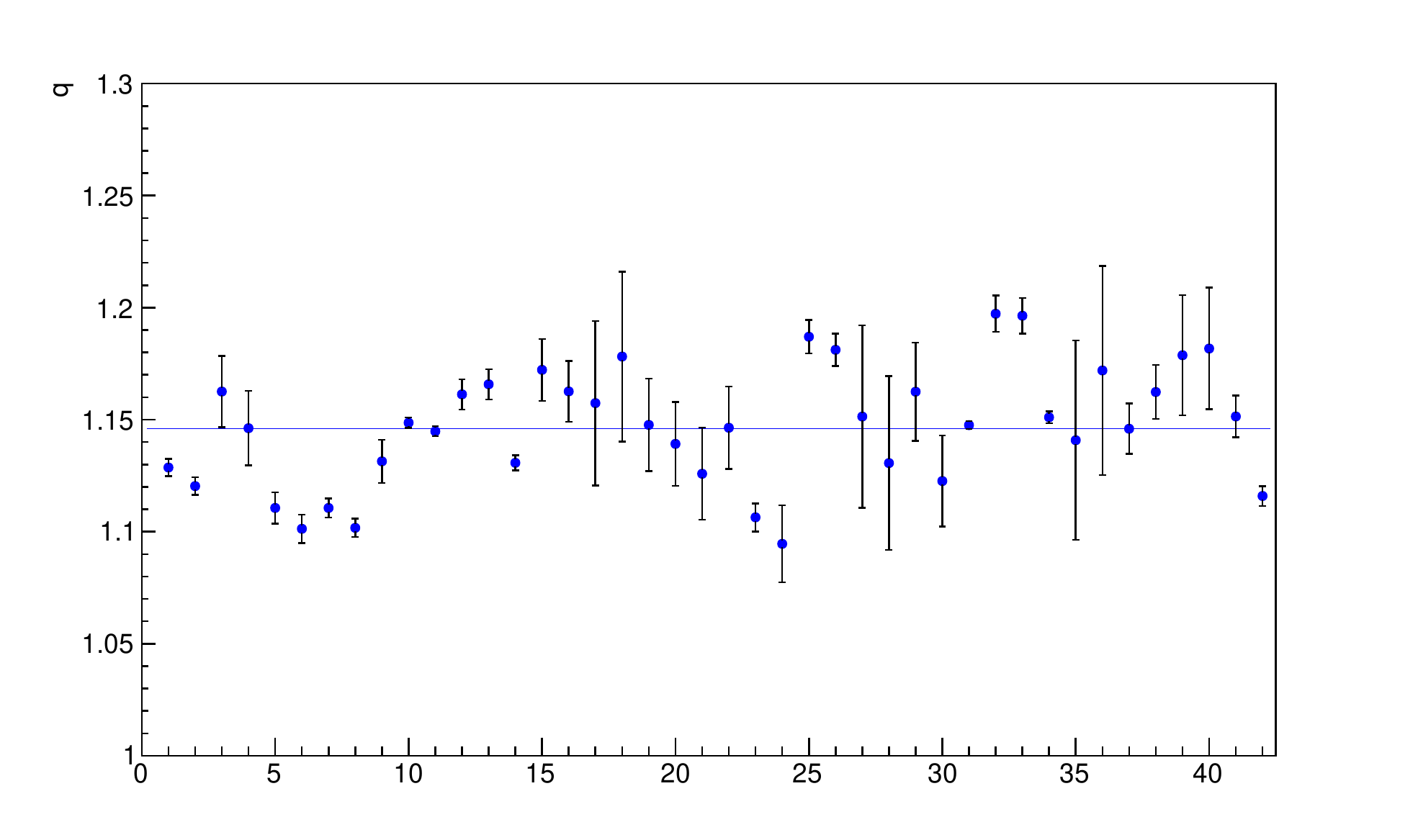}
    }
    \caption{Results for the temperature, $T$ and for the entropic index, $q$, for all the sets of data analysed in this work.
	  The numbering on the abscissa corresponds to the entries in Table 1.}
    \label{fig:Tq}
 \end{figure}

In Fig. \ref{fig:ExAjustedN} we show a typical result for the fitting of Eq.~\ref{fittingformula} to 
measured $p_T$-distribution. With the fittings for all experimental data used here we obtain the values 
for $T$ and $q$ that are plotted in Fig. \ref{fig:Tq}, where we observe a distribution of both temperature and entropic 
index around a constant value.

The results obtained here are in agreement with previous analysis both qualitatively and quantitatively. 
Therefore we can determine the mean values from the analyses~\cite{CleymansWorku, Lucas, Sena1, Sena2}, 
obtaining $q = 1.146 \pm 0.004$ and $T = 68 \pm 5$ MeV. With these values all the thermodynamical aspects of the 
hot fireball are determined.

\cleardoublepage
\newpage

\section{Rapidity Distribution}

In the Bjorken model~\cite{Bjorken} a plateau in the central region of the rapidity distribution  
exists reflecting the invariance of the fireball under Lorentz transformation for boosts 
in the beam direction. Experimental results for high energy collisions, however, have shown that such a plateau 
does not exist, data showing two small peaks symmetrically located near the central rapidity region.

A systematic analysis performed in Ref.~\cite{A_Kumar} shows that color-glass-condensate (CGC) approaches give a 
good description of the results at the central region but fail to reproduce the data at large values of $|y|$. 
Here we use a model for describing the rapidity distribution in pp collision using the 
non-extensive distribution. A similar method was already used in Refs~\cite{CB, Cleymans2008} in terms of 
Boltzmann-Gibbs statistics but the present analysis keeps some differences with respect to the previous one, as 
will be detailed below. 

Suppose that the fireball is moving in the laboratory frame in the beam direction with rapidity $y_f$. According 
to Eq.~\ref{eq:doublediff} the yield of secondaries from the decay of this fireball is given by
 \begin{equation}
 \frac{d^2N}{p_T\,dp_T\,dy}= gV \frac{m_T \cosh y'}{(2\pi)^2}\bigg[1+(q-1)\frac{m_T \cosh y'-\mu}{T} \bigg]^{-\frac{q}{q-1}}
 \label{eq:6}
\end{equation}
where $y = y' + y_f$.

From now on we assume $\mu = 0$ since we restrict ourselves to high energy pp collisions, and  adopt a model 
for the fireball longitudinal expansion which is similar, but not identical, to the Bjorken 
scenario~\cite{Bjorken}. In the present model the system  the basic 
fireballs have rapidity distributed according to the function $\nu(y_f)$ so that
\begin{equation}
 \frac{d^2N}{p_T\,dp_T\,dy}= \frac{N}{A}\int_{-\infty}^\infty \nu(y_f) \times  \frac{m_T \, \cosh(y-y_f)}{(2\pi)^2} \left[1+(q-1)\frac{m_T \cosh(y-y_f)}{T} \right]^{-\frac{q}{q-1}} dy_f
 \label{eq:7}
\end{equation}
where $\nu(y_f)dy_f$ is the number of fireballs with rapidity between $y_f$ and $y_f + dy_f$ , $N$ is the
fully integrated particle multiplicity and $A$ is a normalization constant ensuring that
\begin{equation}
 \frac{1}{N}\int_{-\infty}^{\infty} \int_0^{\infty} \frac{d^2N}{dp_T\,dy} dp_T\,dy=1\,.
\end{equation}

It is possible to integrate Eq.~\ref{eq:7} over the transverse momentum $p_T$, resulting in:
\begin{small}
\begin{align}
\frac{1}{N}\frac{dN}{dy} &=& \frac{1}{A} \int_{-\infty}^{\infty} \nu(y_f) 
\left[ T^3 \sech^2(y-y_f) \left(\frac{(q-1)\cosh (y-y_f)}{T}\right)^{\frac{2 q-3}{q-1}} (m_0 (q-1)+T \sech(y-y_f))\right]
\nonumber \\
 & & \times \left[\frac{\left(m_0+\frac{T \sech(y-y_f)}{q-1}\right)^{-\frac{q}{q-1}}\left(- m_0^2 (q-2)+2 m_0 T \sech(y-y_f)+2 T^2\sech^2(y-y_f)\right)}{4 \pi ^2 (q-2) (q-1)^3(2 q-3)}\right] dy_f\,.
\label{eq:fitformula}
\end{align}
\end{small}
The equation above gives the rapidity distribution of particles produced in pp collisions and is the basis
for describing the available experimental results at different beam energies.

For completeness we still need the distribution function $\nu(y_f)$. 
Based on the shape of the experimental distributions we choose the following  ansatz:
\begin{equation}
 \nu(y_f)= G_{q'}(y_0,\sigma;y_f) + G_{q'}(-y_0,\sigma;y_f)\,, \label{eq:yfdist}
\end{equation}
where
\begin{eqnarray}
G_q(y_0,\sigma;y_f) = \frac{1}{\sqrt{2 \pi } \sigma} e_q\left(-\frac{(y_f - y_0)^2}{2\sigma^2}\right)\,,
\label{eq:qgauss}
\end{eqnarray}
and $e_q(x)$ is the q-exponential function defined as
\begin{align}
 e_q(x) & \equiv [1 - (q-1)x]^{-1/(q-1)} \,.
\label{eq:qexp}
\end{align}
In Eq.~\ref{eq:qgauss} $y_0$ and $\sigma$ are respectively the peak position and the width of the 
q-Gaussian function and are considered here as free parameters to be adjusted in a systematic analysis of 
experimental data. Eq.~\ref{eq:yfdist} assumes that the main final state  in 
the collision is composed of two  fireballs moving in the beam direction with 
opposite rapidities $y_f$,  each of them being composed by fireballs with rapidity-gaps with respect to the 
corresponding cluster that are distributed according to a q-Gaussian function.

Notice that $q'$ in Eq.~\ref{eq:yfdist} does not need to be equal to $q$ and in the present work we perform the 
analysis of experimental data with two different assumptions: $q'=q$ and $q'=1$, the last assumption corresponding 
to a Gaussian distribution of the rapidity-gap of the fireballs in each of the moving cluster. 
In addition, we assume that the fireballs are described by the non-extensive self-consistent thermodynamics with 
the limiting temperature $T$ and the characteristic entropic index, $q$, that were found in the 
analysis of $p_T$-distributions and hadronic spectrum, as described above. The values adopted here 
are $T=$68~MeV and $q=$1.146. For simplicity we will adopt $m_0\,=\,$ 139.59~MeV, since most of the charged hadrons 
produced at high energy collisions are pions. 
The set of experimental data used for the rapidity analysis is shown in Table~\ref{table:rapidity}.

\begin{table}[!ht]\centering
\caption{ }
\scalebox{0.85}{
\begin{tabular}{c c c c} \toprule
 $\sqrt{s}\,(GeV)$& Exp.    & Range  &     Ref.          \\ \midrule
  53              & UA5     & $ |y| \leq 3.1$ & \cite{rapidity53GeV}   \\
  200             & UA5     & $ |y| \leq 4.6$ & \cite{rapidity53GeV}   \\
  200             & PHOBOS  & $ |y| \leq 5.3$ & \cite{rapidity200e410GeV}   \\
  410             & PHOBOS  & $ |y| \leq 5.3$ & \cite{rapidity200e410GeV}   \\
  546             & UA5     & $ |y| \leq 4.8$ & \cite{rapidity53GeV}   \\
  630             & UA5     & $ |y| \leq 5.5$ & \cite{rapidity630GeV}   \\
  900             & UA5     & $ |y| \leq 4.6$ & \cite{rapidity53GeV}   \\
  900             & CMS     & $ |y| \leq 2.5$ & \cite{rapidity900e2360GeV}   \\
  1800            & CDF     & $ |y| \leq 3.5 $ & \cite{rapidity1800GeV}   \\
  2360            & CMS     & $ |y| \leq 2.5$ & \cite{rapidity900e2360GeV}   \\
  7000            & CMS     & $ |y| \leq 2.5$ & \cite{rapidity7TeV}   \\ \bottomrule
\end{tabular} }\label{table:rapidity}
\end{table}

The main difference between the analysis performed in the present work and that performed in 
Refs.~\cite{CB, Cleymans2008} is, of course, the use of Tsallis statistics in the thermodynamical 
description of the fireball however other differences must be noticed. While in~\cite{CB}  both 
temperature and chemical potential could depend on the fireball rapidity, here we adopt constant 
temperature at the limiting value predicted theoretically and null chemical potential regardless 
of the fireball rapidity. 
This approach is justified by the fact that our analysis focuses on charged hadrons 
distributions produced  at very high 
beam energies ($>$ 500 GeV), which are dominated by pion production. This assumption can be improved 
with the availability of rapidity distributions for identified hadrons~\cite{stiles,CB,Cleymans2008}.

 \begin{figure}[!ht]
    \centering
    \includegraphics[scale=0.9]{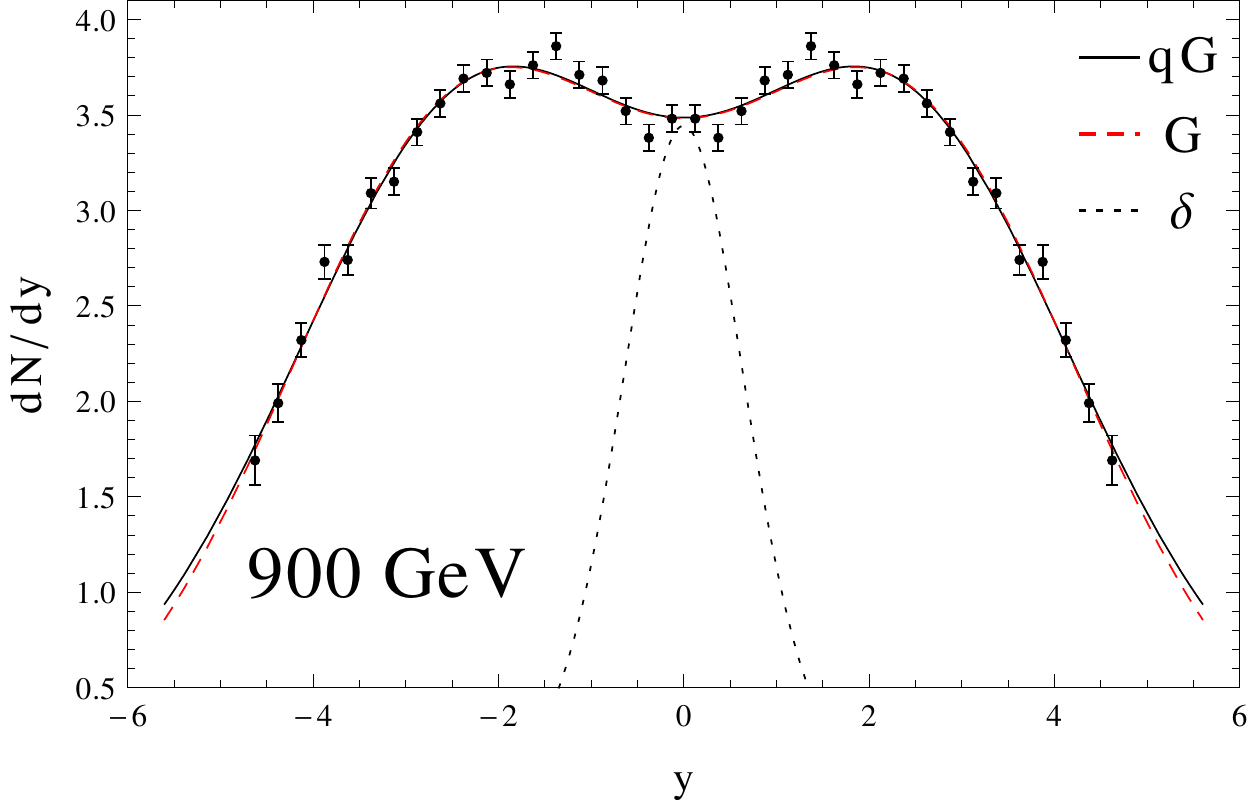}
    \caption{Fit using Eq.~\ref{eq:fitformula} to experimental data for rapidity distribution from pp 
    collision at 900 GeV from the UA5 collaboration~\cite{rapidity53GeV}. 
    Data for $y<0$ are obtained symmetrically from $y>0$.  The full line represents the best fit 
    using the q-Gaussian function ($q'=q$) and the dashed line corresponds to the best fit using 
    the Gaussian function ($q'=1$) in Eq.~\ref{eq:yfdist}. 
    The dotted line shows the results when a  delta function is used for the 
    fireball rapidity distribution, i.e., $\nu(y_f)=\delta(y_f)$.}
    \label{fig:typicalfit}
 \end{figure}

\begin{figure}[!ht]
   \centering
   \subfigure[]{
   \includegraphics[scale=0.8]{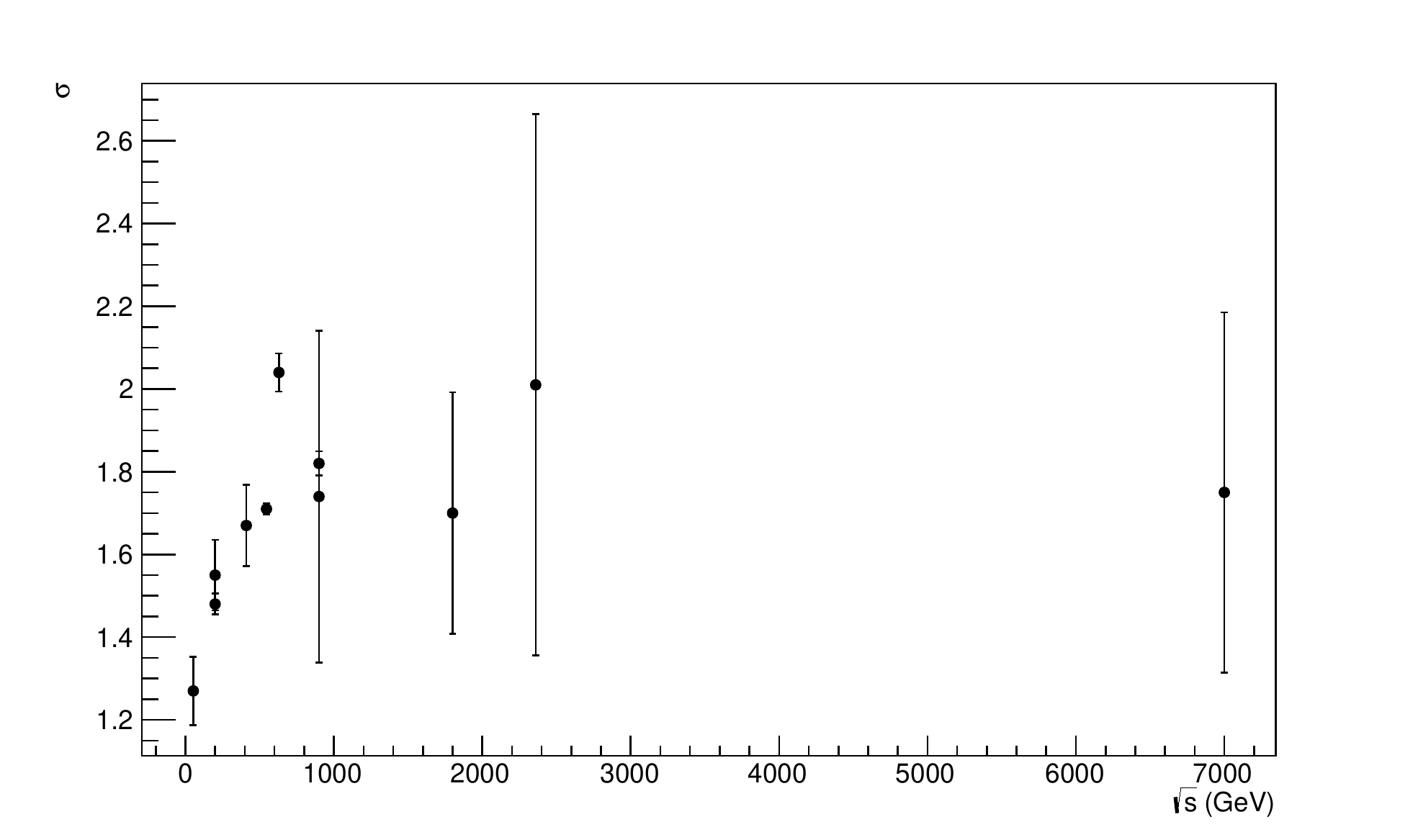}
   }
   \subfigure[]{
   \includegraphics[scale=0.8]{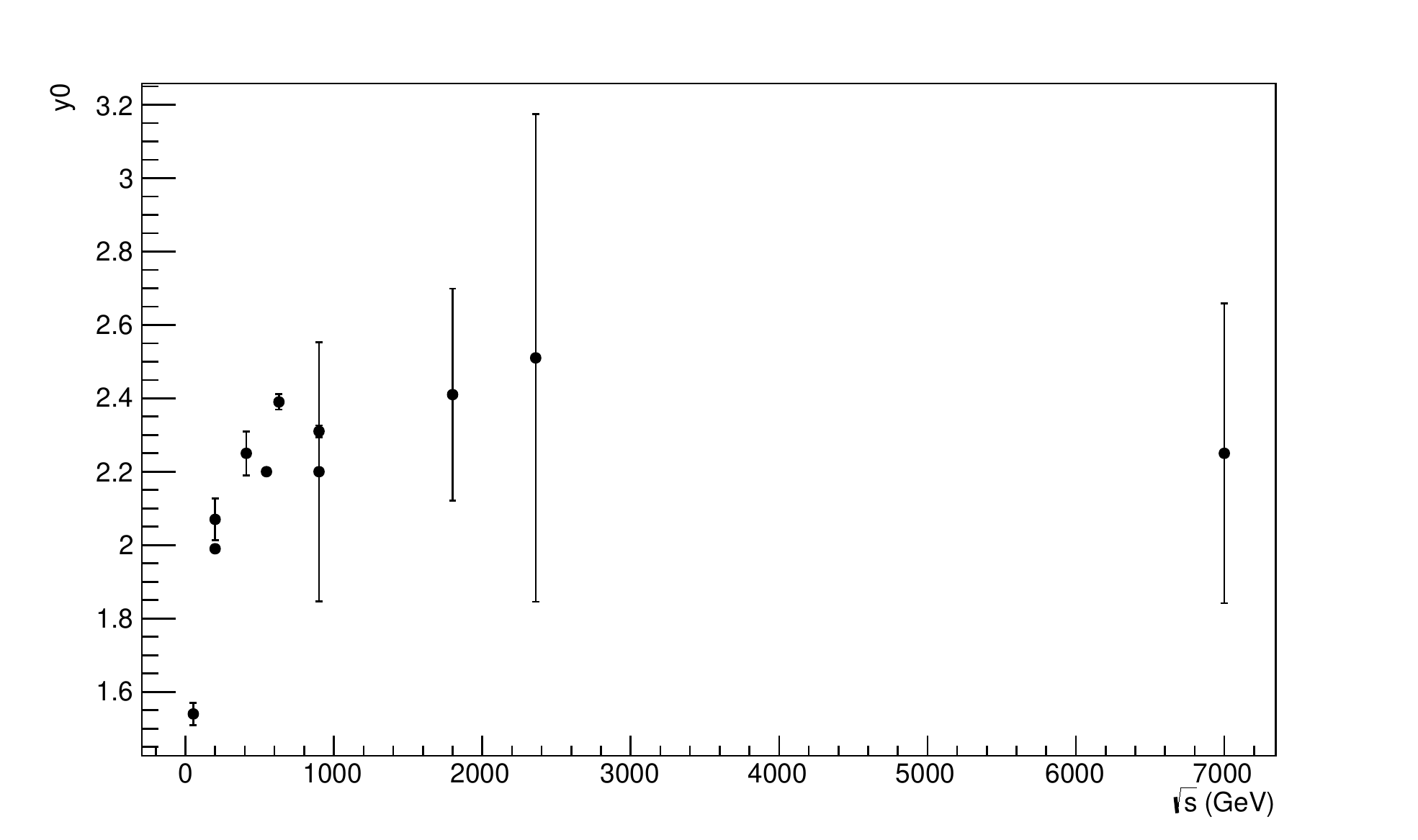}
   }
   \caption{Values for the parameters $\sigma$ and $y_0$ in Eq.~\ref{eq:fitformula}  as a function of the collision energy.}
   \label{fig:1}
\end{figure}

In Fig.~\ref{fig:1} we show the results of fits to the rapidity distributions of particles produced in pp collisions 
from several experiments with collision energies ranging from 50~GeV up to 7~TeV. 
The set of experimental data used in the complete analysis is shown in Table \ref{table:rapidity}. 
We observe a good fit of formula~\ref{eq:fitformula} to the experimental data in the whole range of rapidity available, which 
covers almost the entire rapidity region. These results and the others presented in Appendix A shows that a 
consistent description of the experimental data can be obtained with the model described above. 
Notice that the only free parameters in the fitting procedure are $\sigma$ and $y_0$ appearing in 
Eq.~\ref{eq:yfdist}. From the analysis performed on all the sets of experimental data in 
Table \ref{table:rapidity} we get the values for those parameters that are presented in Fig.~\ref{fig:typicalfit}.
 
For both quantities we observe a sharp increase up to collision energies around 500~GeV. 
The uncertainties above 
this energy are also larger due to the fact that the range of rapidity available is smaller. 
The striking feature of these results is the fact that both the peak position $y_0$ and the width $\sigma$ of the 
q-Gaussian remain within the same range i.e. are approximately constant for collision energies from 
500~GeV to 7~TeV, while the beam rapidity 
varies by a factor 2 in this range of collision energies. These results allow us to conclude that after the two original 
protons collide two clusters of fireballs are formed and they move  parallel to the beam direction with opposite 
rapidities given by $|y_0|= 2.3$. Each cluster is formed by fireballs that move with respect to the center of each 
cluster in the beam direction with rapidity distribution that is described by a q-Gaussian function with 
width $\sigma\approx 1.8$.  
We will assume in the rest of this paper that the values of $y_0$ and $\sigma$ will not change
drastically for collision energies up the 14 TeV. 
Of course, it is expected that the position of $y_0$ will increase at much higher beam energies.

Below $\approx$~550~GeV the scaling observed above disappears, as can be seen in Fig.~\ref{fig:1}. 
It is possible that at 
these collision energies the temperature achieved by the system is significantly lower than 
the limiting temperature, $T$, hence
more studies on the direction of verifying the role played e.g. by the chemical potential at low energies 
must be carried out, as performed in~\cite{stiles,CB,Cleymans2008}.

When we use the Gaussian function instead of the q-Gaussian for the rapidity distribution of the fireballs, i.e.,
\begin{equation}
 \nu(y_f)= G(y_0,\sigma;y_f) + G(-y_0,\sigma;y_f)\,, \label{eq:yfdistg}
\end{equation}
with $G(y_0,\sigma;y_f)$ being the Gaussian function with peak position at $y_0$ and width $\sigma$, the results of 
the fitting procedure Eq.~\ref{eq:fitformula} to the experimental data are very similar to those obtained 
with the q-Gaussian function, as shown in Fig.~\ref{fig:typicalfit} with dashed line. 
The best fit values for the parameters $y_0$ and $\sigma$ are also practically 
the same and they are presented in Fig.~\ref{fig:1} with open symbols.

\section{Energy Dependence of Multiplicity}


Once the parameters $T$, $q$, $y_0$ and $\sigma$ are determined, 
if we know the function $N(\sqrt{s})$ giving the 
multiplicity as a function of energy we can determine the $p_T$ and rapidity-distributions of secondaries produced in 
pp collisions at any collision energy sufficiently high to allow the assumption that those parameters are 
independent of 
beam energy and particle properties. Therefore now we investigate how the function $N(\sqrt{s})$ behaves.

The energy carried out by all particles with mass $m$ is determined by
\begin{equation}
 E(m) =  \frac{gV}{(2\pi)^2}  \rho(m) \int_{-\infty}^\infty \int_{0}^\infty \varepsilon(p_T,y) \frac{d^2N}{dp_T\,dy} dp_T\,dy\,,
 \label{eq:13}
\end{equation}
where
\begin{equation}
 \varepsilon(p_T,y)= m_T \cosh(y)
\end{equation}

The total energy carried by baryons is 
\begin{equation}
 E =  \int_0^\infty \rho(m) E(m) dm
 \label{eq:14}
\end{equation}
where $\rho(m)$ is the baryon mass spectrum. The function $\rho(m)$ was determined by the non-extensive self-consistent theory~\cite{Deppman} and is given by
\begin{equation}
 \rho(m) = \frac{\gamma}{m^{5/2}}\left[ 1 + \frac{(q - 1)m}{T} \right]^{\frac{1}{q-1}} 
 \label{massspectrum}
\end{equation}
In Ref.~\cite{Lucas} it was shown that this formula  reproduces correctly the observed hadronic mass spectrum
up to a mass of approximately 2 GeV. 

Substituting Eq.~\ref{eq:7} in Eqs.~\ref{eq:13} and~\ref{eq:14} we obtain a relation between the total baryonic energy 
and the multiplicity that is given by
\begin{equation}
 E=N(E)\frac{1}{A} \int_0^\infty \rho(m) \int_{-\infty}^\infty  \int_{0}^\infty \varepsilon(p_T,y) \frac{d^2N}{dp_T\,dy} dp_T\,dy\,dm\,.
\end{equation}
Notice that for sufficiently high energy, when $T$, $q$, $\sigma$ and $y_o$ do not depend on the collision energy, the 
only term which can depend on the total energy is the multiplicity, $N(E)$. At the limiting temperature the calculations 
are difficult to be performed because of the singularity in the partition function corresponding to the phase transition. 
Therefore we use a sum over a subset of the particles produced in high energy collisions, so that
\begin{equation}
 N(E)=K E\,, \label{eq:ME}
\end{equation}
where
\begin{equation}
 K=\bigg[ \frac{1}{A} \sum_{i=1}^n \int_{-\infty}^\infty  \int_{0}^\infty \varepsilon(p_T,y) \frac{d^2N}{dp_T\,dy}(m_i) dp_T\,dy\bigg]^{-1} \label{eq:K}
\end{equation}
where the sum is over all the particles indicated in Table~\ref{table:particles}, $m_i$ is the particle mass, and 
\begin{equation}
 \frac{d^2N}{dp_T\,dy}(m_i)
\end{equation}
is the double differential yield given by Eq.~\ref{eq:fitformula} when the mass $m_i$ is used.

\begin{table}[!ht]\centering
\caption{ }
\scalebox{0.85}{
\begin{tabular}{c c } \toprule
  Particle     & Mass (GeV/c$^2$)     \\   \midrule
  $\pi^{\pm}$           & 0.140             \\
  $K^{\pm}$             & 0.494              \\
  $\rho$                & 0.770                \\
  $K^*$                 & 0.892                \\
  $p$                   & 0.938              \\
  $\Sigma^+$            & 1.189               \\
  $\Sigma^-$            & 1.197               \\
  $\Delta$              & 1.232                \\  
  $\Xi^-$               & 1.321               \\
  $\Sigma^*$            & 1.385                \\
  $\Xi^*$               & 1.533                \\
  $\Omega^-$            & 1.672                \\ 
  $D^{\pm}$             & 1.869                \\
  $F^{\pm}$             & 1.971                \\
  $D^*$                 & 2.010                \\  
  $\Lambda^+_c$         & 2.281                \\  
  $B^{\pm}$             & 5.271                \\   \bottomrule
\end{tabular} }\label{table:particles}
\end{table}

\begin{figure}[!ht]
   \centering
   \includegraphics[scale=0.8]{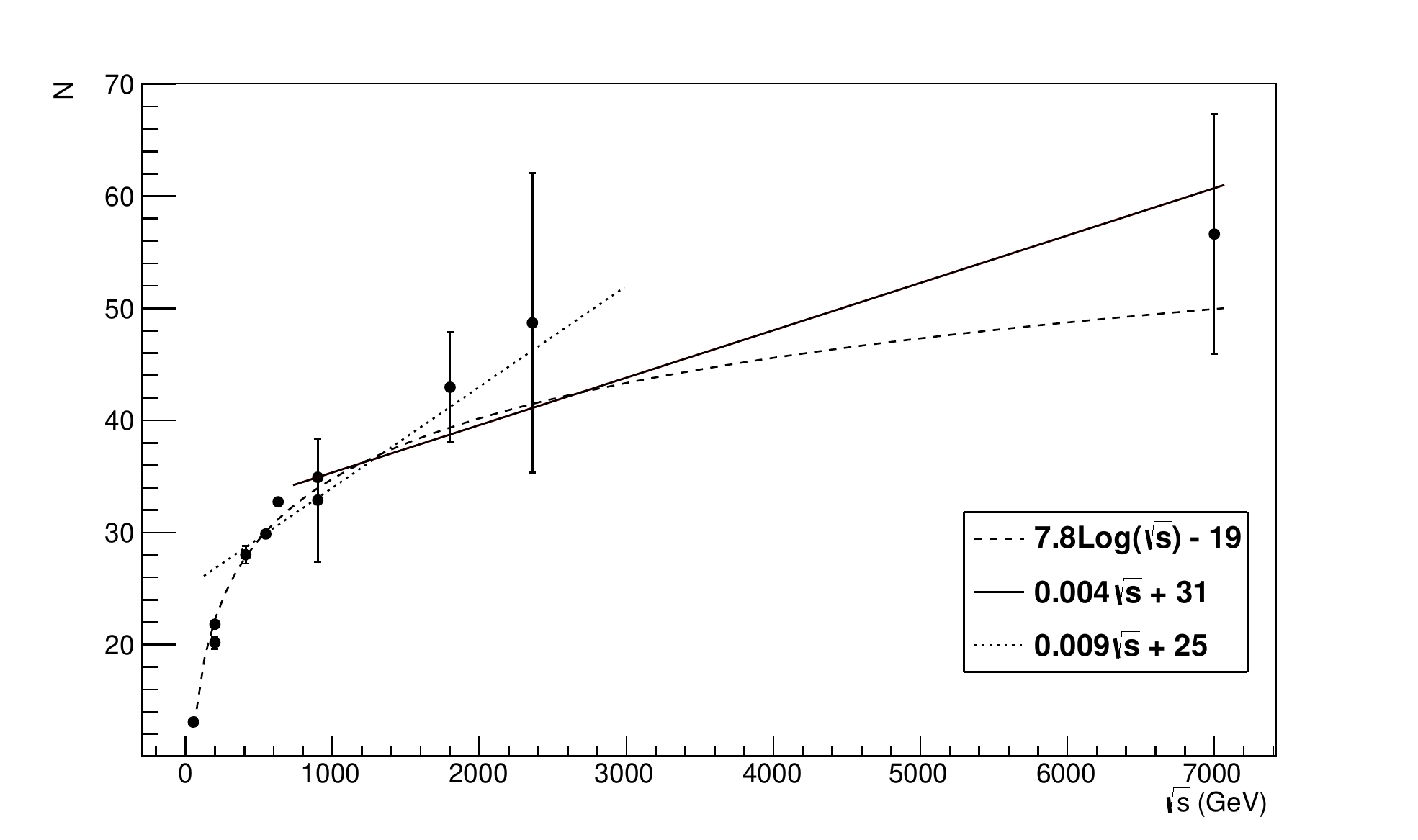}
   \caption{The particle multiplicity $N$ as a function of $\sqrt{s}$. The solid line is a linear fit to the data for $\sqrt{s}>$~800~GeV. The dashed line represents the  logarithm function. The dotted line is the multiplicity as a function of energy when the coefficient $K$ is used.}
   \label{fig:AE}
\end{figure}

 With Eq.~\ref{eq:K} and using all particles listed in Table~\ref{table:particles} we get $K=0.009$~GeV$^{-1}$. This value is above the best fitting value $K=0.004$~GeV$^{-1}$, showing that the more massive particles not included in our calculation give a significant contribution to the total energy. In Fig.~\ref{fig:AE} we show the linear variation of multiplicity using the calculated value for $K$ as a dashed line.
 
In Fig.~\ref{fig:AE} we present the multiplicity $N$ obtained from the fittings of Eq.~\ref{eq:fitformula} to the rapidity distribution data using the q-Gaussian function for the fireball rapidity distribution (black symbols). We observe that after a fast increase for collision energies up to $\sim$~500~GeV the multiplicity increases more slower at higher energies. As already discussed, for low energies we cannot ensure that $T$ is constant and that the chemical potential is null, therefore we will not deepen our discussion for the results at those energies. Nonetheless we fitted the multiplicity data with two different fitting formulas: one linear for energies above 500~GeV and the other logarithm for the full rapidity range. The results are plotted in Fig.~\ref{fig:AE}.

We observe that the linear fit has a coefficient very different from $K$ calculated above. Indeed, 
the curve one would obtain with this coefficient is the dotted line also shown in that figure, which gives a consistent description of the 
multiplicity in the energy range from 500~GeV up to 2500~GeV but would not describe correctly the multiplicity at 7~TeV. 
This happens because the coefficient $K$ consider only the contribution from the hadrons listed in Table~\ref{table:particles}, which results in an overestimation of the total energy carried by all hadrons according to Eq.~\ref{eq:ME}.

The problem is that as the collision energy increases, small variations in the temperature, which is tending to the limiting temperature, will result in a sharp increase of the number of massive hadrons, as described by $\rho(m)$, and these particles contribute significantly to the total energy even if they have small contributions to the total multiplicity. Therefore the linear fit in Fig.~\ref{fig:AE} can be estimated as an upper limit to the multiplicity for energies above 7~TeV. The logarithm fit, on the other hand, could be considered as the lower limit, since it systematically underestimate the charged hadrons multiplicities for collision energies above 2~TeV.

As examples of application of the present model we now calculate the $p_T$- and $y$-distributions for charged particles at collision energy $E = 13$~TeV, which is the expected energy at the LHC experiments.

\begin{figure}[!ht]
   \centering
   \includegraphics[scale=0.9]{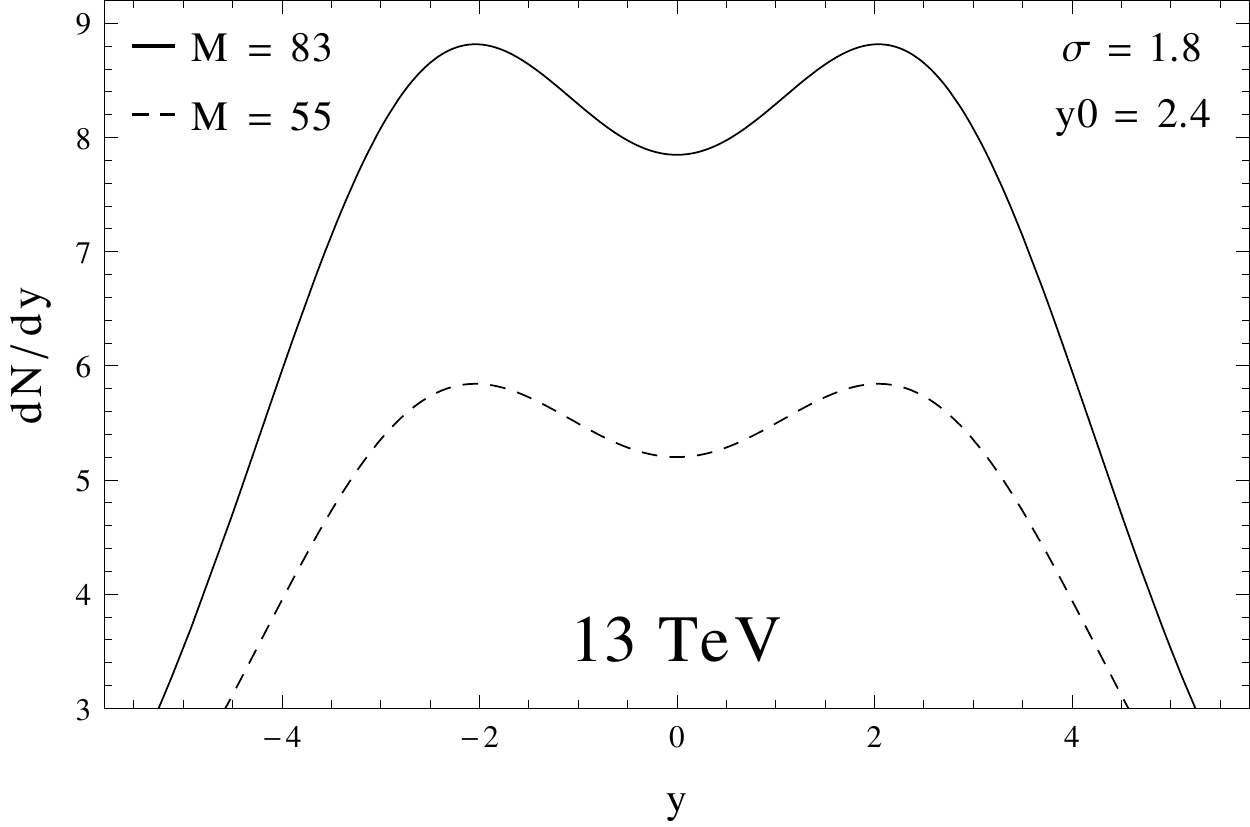}
   \caption{The charged hadrons rapidity-distribution predicted for 13~GeV pp collisions.  
	   The full line corresponds to the multiplicity calculated through the linear extrapolation and the  
	   dashed line corresponds to the multiplicity obtained with the 
   logarithm extrapolation, as given in Fig.~\ref{fig:AE}. } \label{predy}
\end{figure}

\begin{figure}[!ht]
   \centering
   \includegraphics[scale=0.9]{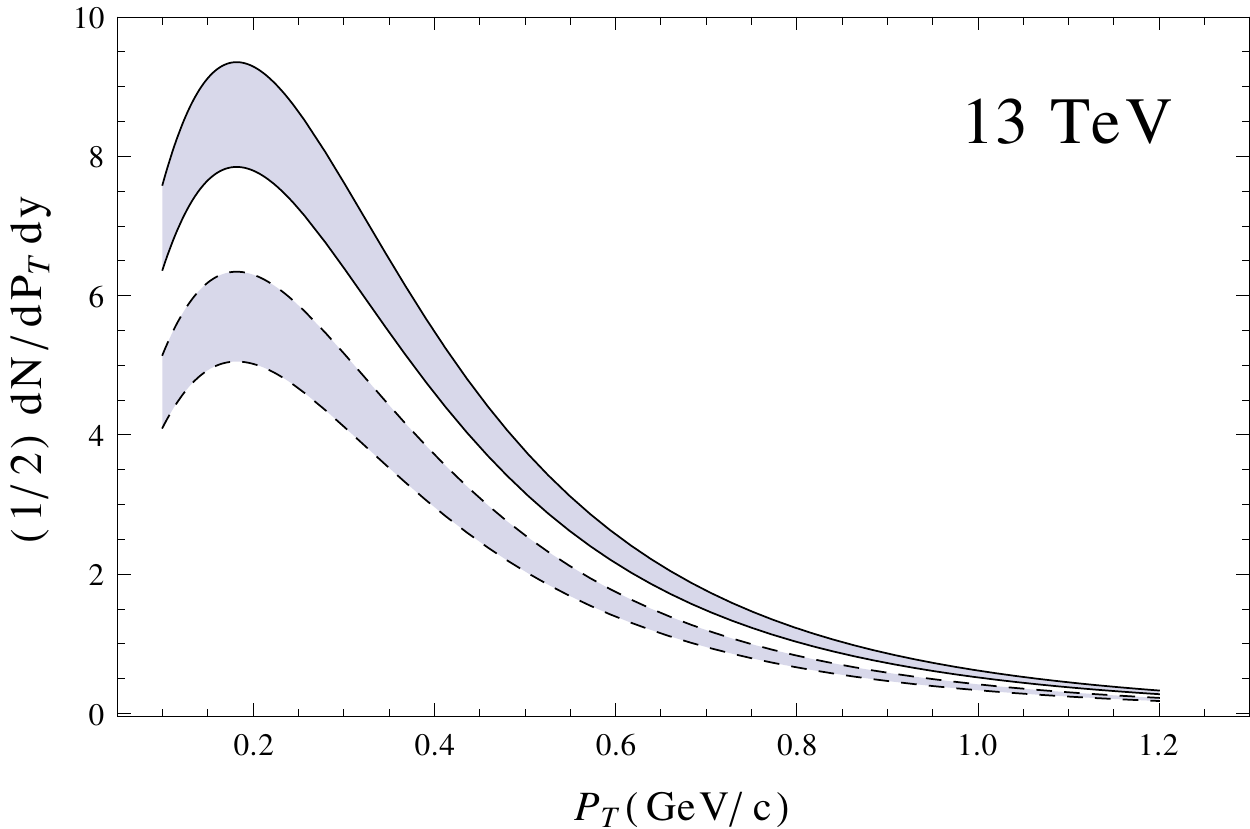}
   \caption{The charged hadrons $p_T$-distribution predicted for 13~GeV pp collisions. The shaded regions show the 
   expected distribution considering the uncertainties in the parameters $T$ and $q$. 
   The upper region with full lines corresponds to the multiplicity calculated through the linear extrapolation and 
   the bottom region with dashed lines corresponds to the multiplicity obtained with the 
   logarithm extrapolation, as given in Fig.~\ref{fig:AE}. } \label{predpT}
\end{figure}

The rapidity distribution is shown in Fig.~\ref{predy} using the linear extrapolation (full line), which gives
a total multiplicity $M=83$, and the logarithm extrapolation, which gives $M=55$. 
These results are obtained with q-Gaussian functions for $\nu(y_f)$ with $y_0=2.4$ and $\sigma=1.8$.

The estimates for $p_T$-distribution of charged particles are shown in Fig.~\ref{predpT}. 
The lines indicate the lower and the upper limits of our theoretical calculations using Eq.~\ref{fittingformula} and 
considering the uncertainties on the parameters $q$ and $T$. The uncertainty in the multiplicity is not taken 
into account but we present the results obtained with both extrapolations for $N(E)$, namely, the linear (full lines) and 
the logarithm (dashed lines). The shaded area between the lines show the region where we expect the data to 
be measured at LHC.

\section{Conclusions}

In this work we presented an analysis of the  $p_T$ and rapidity distribution in pp collisions 
assuming that the final state  can be described by two fireballs obeying Tsallis thermodynamics
having a Tsallis temperature $T=68 \pm 5$~MeV and an entropic index $q=1.146 \pm 0.004$ independent of beam energy. 

Our results  on the transverse momentum distributions are in agreement with previous analysis and allows a complete 
characterisation of the thermodynamics of fireballs.

We developed a  model to describe the rapidity distribution with non-extensive thermodynamics which is based 
on the assumption that the hot system formed after pp collisions can be represented as two clusters of fireballs moving in the 
beam direction with rapidities distributed according to a function $\nu(y_f)$. We use a sum of two q-Gaussian functions to 
describe $\nu(y_f)$. It results that for energies above $\sim$ 500~GeV all rapidity distributions can be reproduced with 
constant peak-position and constant width for the q-Gaussian functions.

With these results we estimated  the behaviour of the multiplicity as a function of the collision energy, allowing us 
to present estimates for the future LHC experiments at 13~TeV.

\section{Acknowledgment}

The authors are thankful to R. Martins and to  S.F. Barros for their help with numerical calculations. This work received support from  CNPq, under grant 305639/2010-2 (A.D.), and by FAPESP under grant 2014/21648-1 (L.M).

\cleardoublepage
\newpage

\section*{Appendix}

\begin{figure}[!ht]
\centering
\subfigure[]{
\includegraphics[scale=0.6]{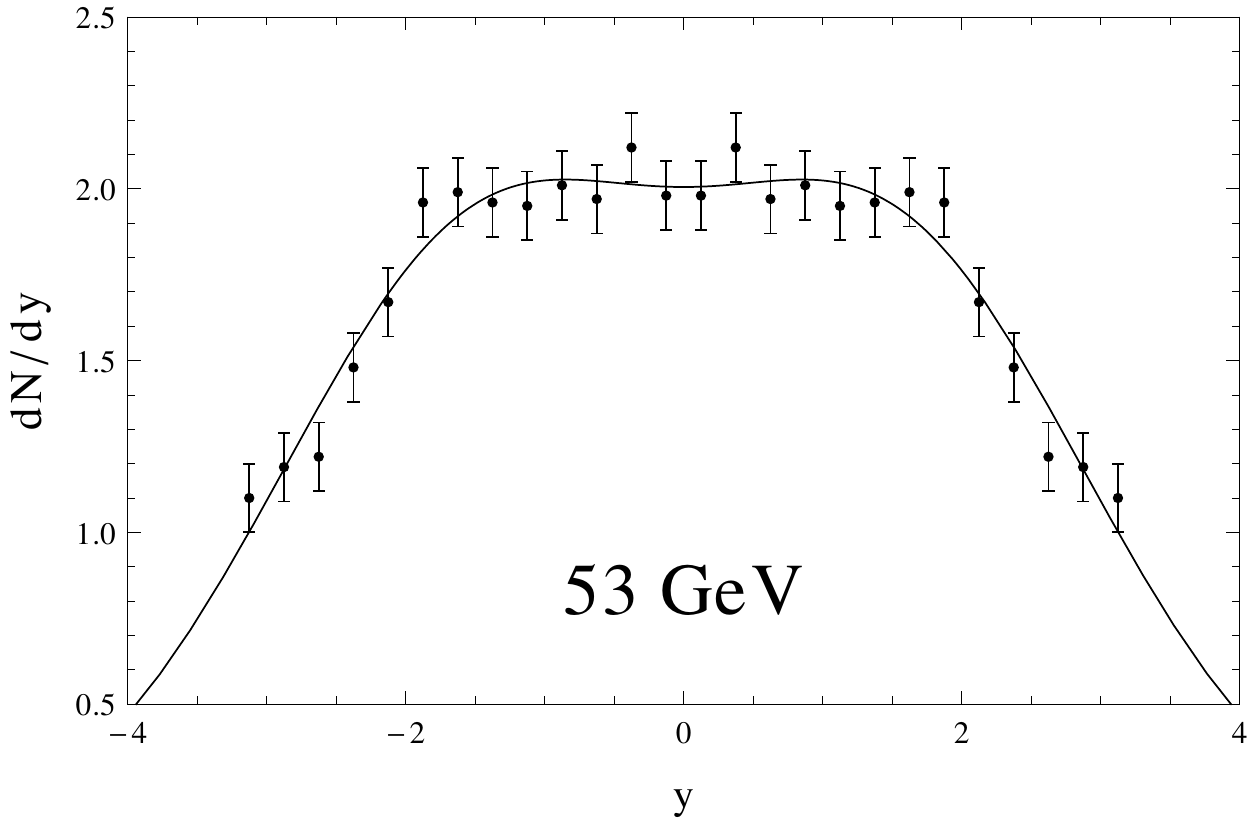}
}
\subfigure[]{
\includegraphics[scale=0.6]{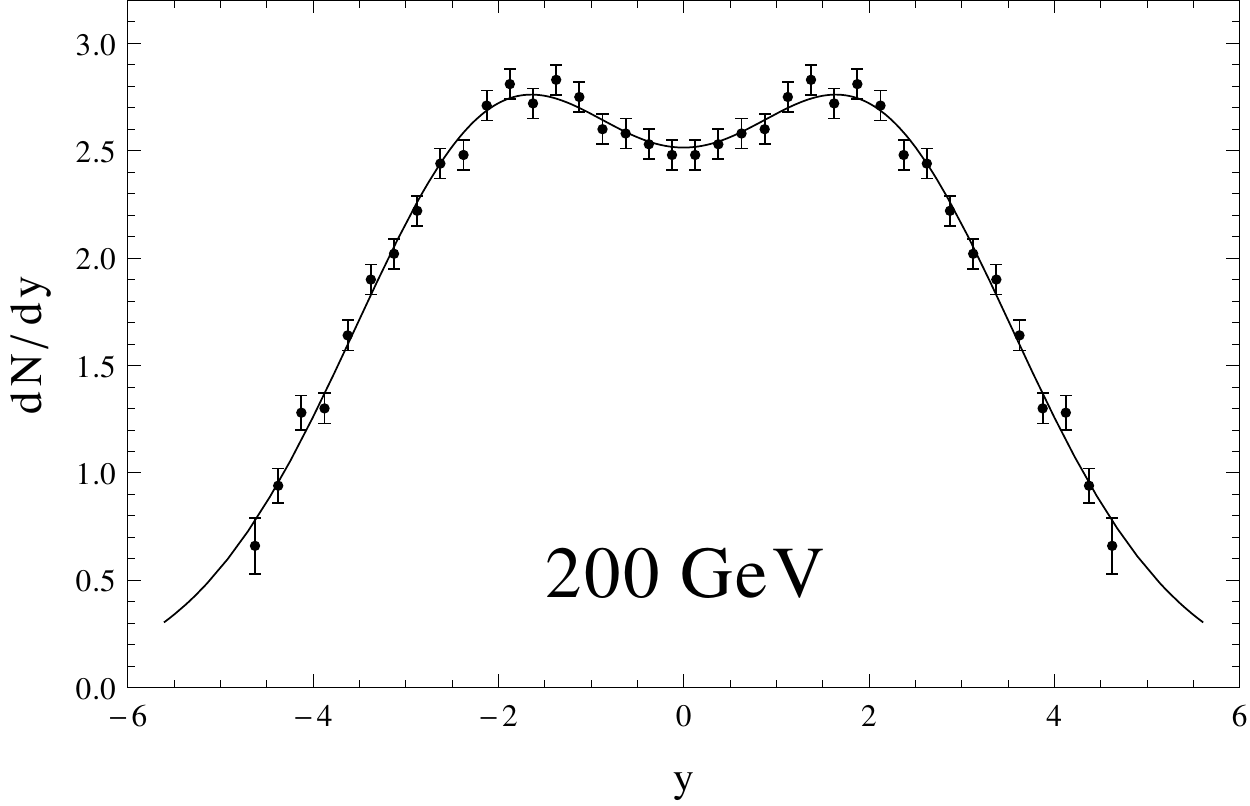}
}
\subfigure[]{
\includegraphics[scale=0.6]{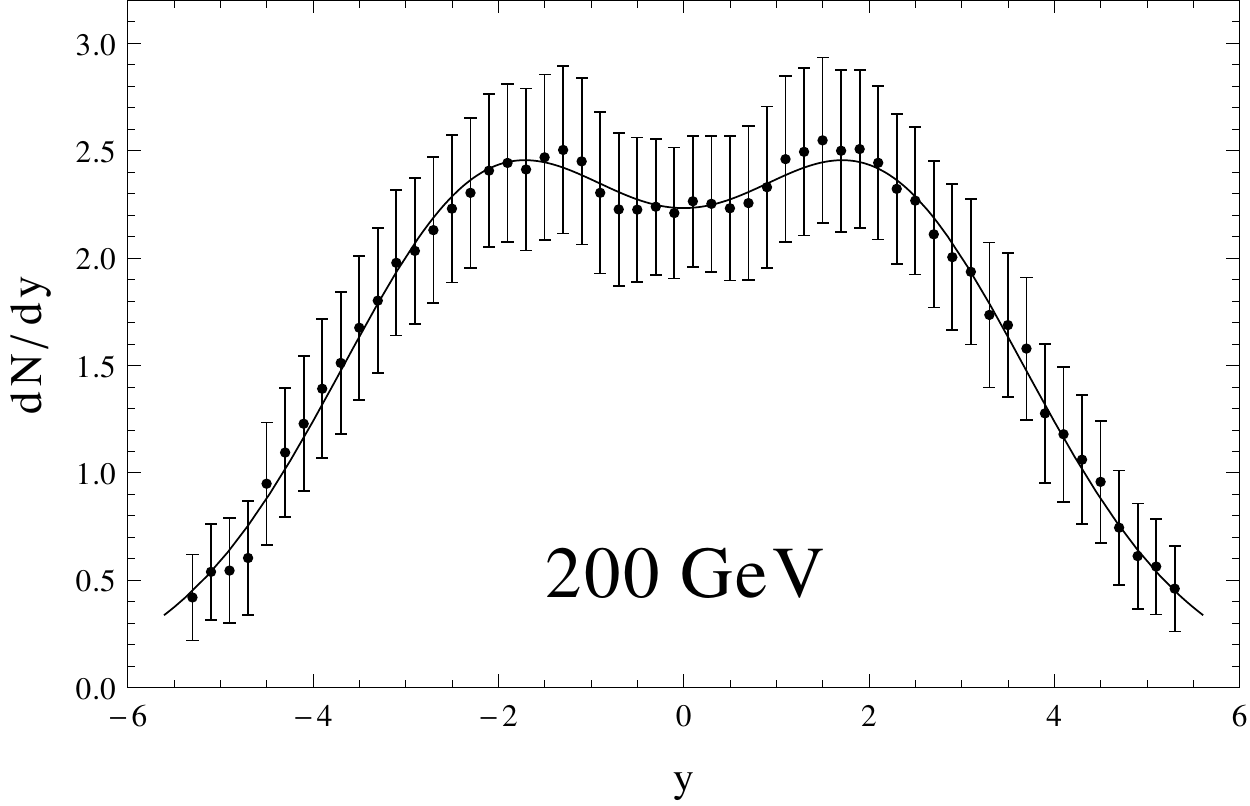}
}
\subfigure[]{
\includegraphics[scale=0.6]{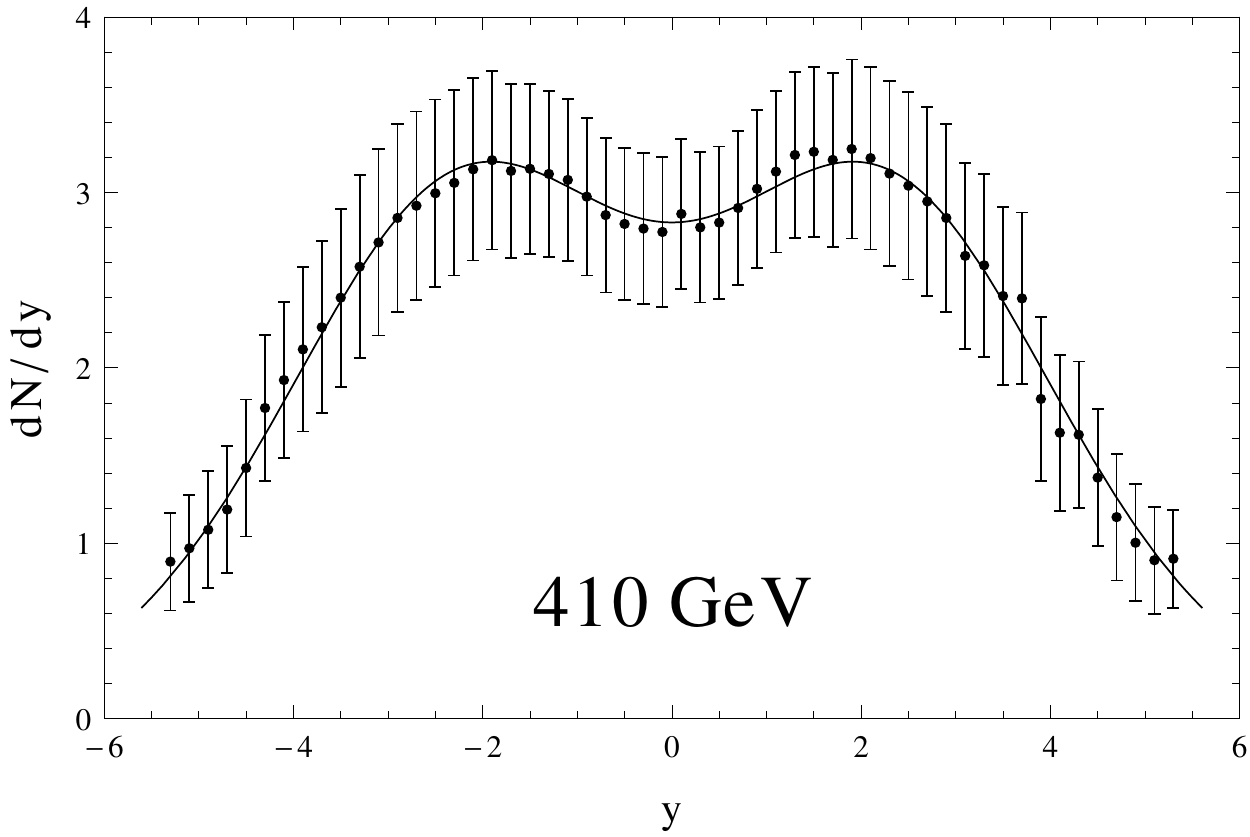}
}
\subfigure[]{
\includegraphics[scale=0.6]{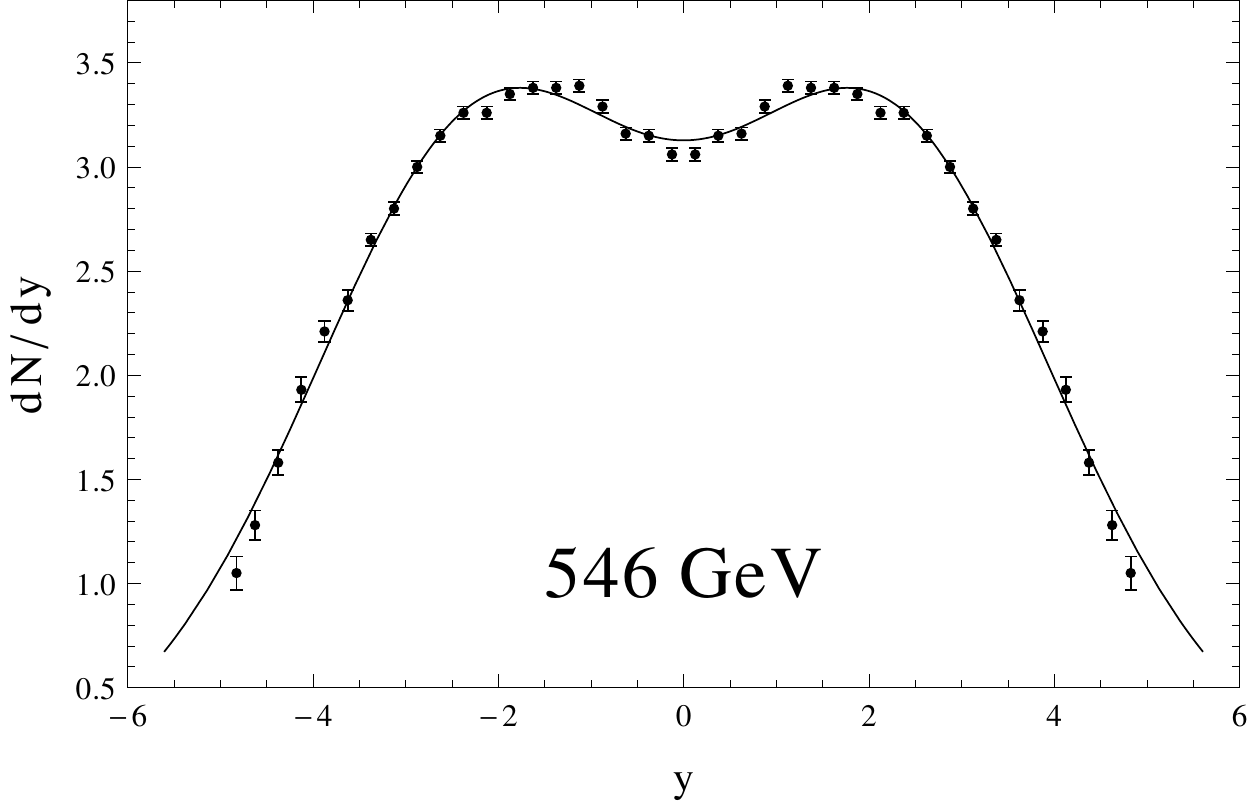} 
}
\caption{Fits to rapidity distributions measured in high-energy p-p collisions (see Table 2) with Eq. \ref{eq:fitformula}. The experimental data are from:  (a) UA5 collaboration~\cite{rapidity53GeV} (b) UA5 collaboration~\cite{rapidity53GeV} (c) PHOBOS collaboration~\cite{rapidity200e410GeV} (d) PHOBOS collaboration~\cite{rapidity200e410GeV} (e) UA5 collaboration~\cite{rapidity53GeV}.}
\label{fig:rapfittings1}
\end{figure}

\begin{figure}[!ht]
\centering
\subfigure[]{
\includegraphics[scale=0.6]{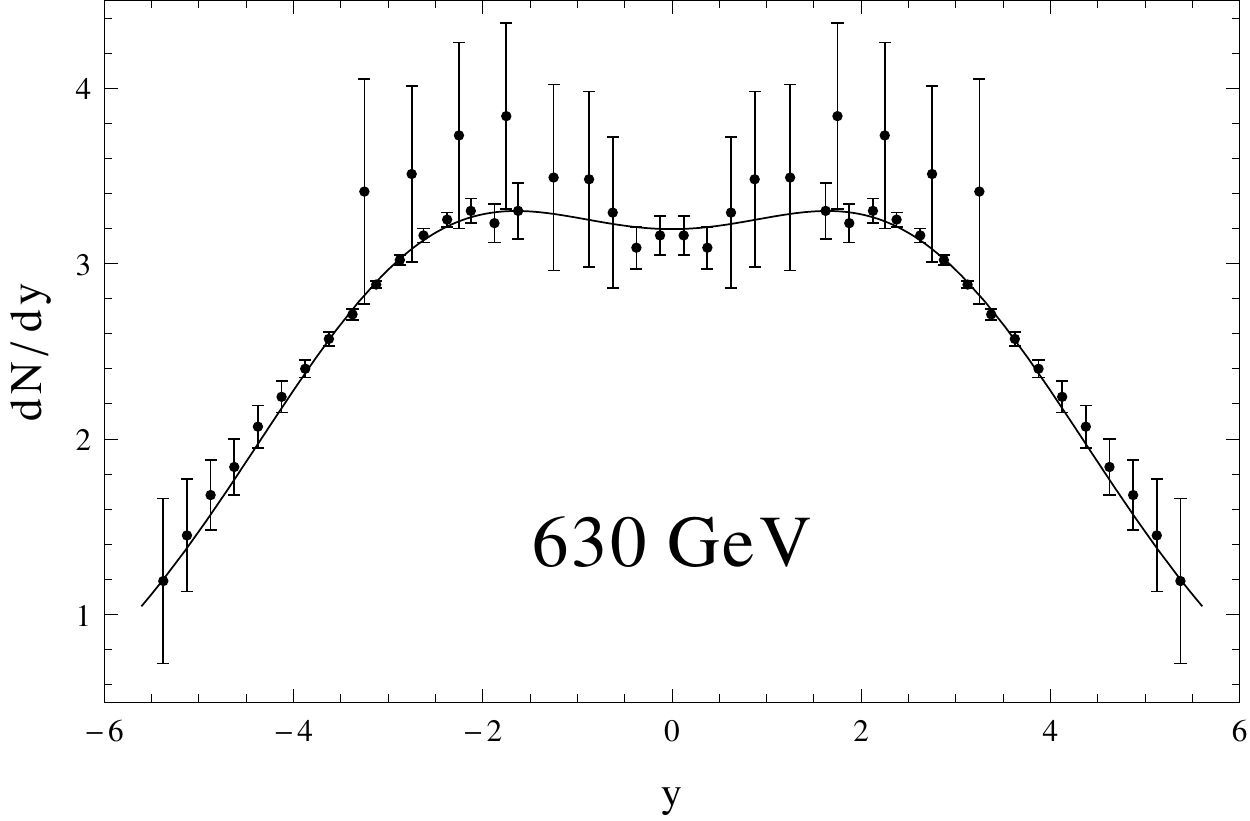}
}
\subfigure[]{
\includegraphics[scale=0.6]{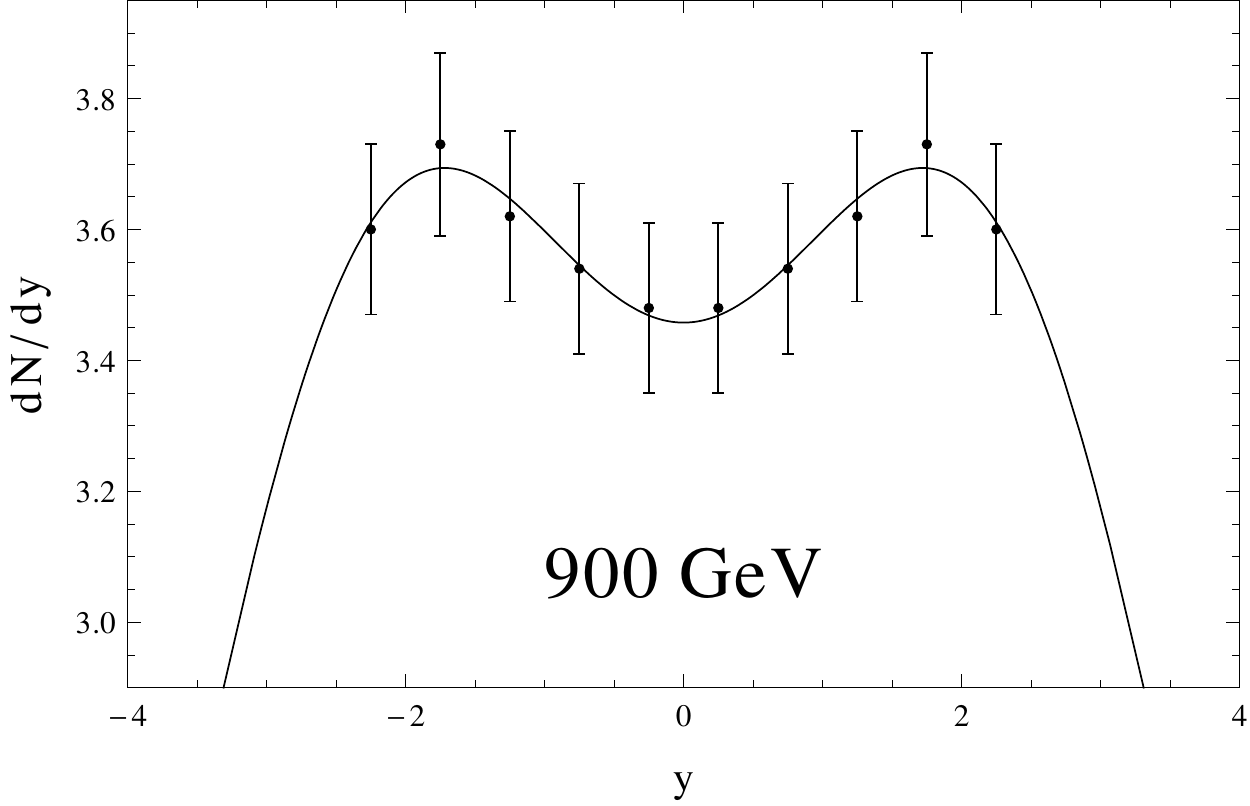} 
}
\subfigure[]{
\includegraphics[scale=0.6]{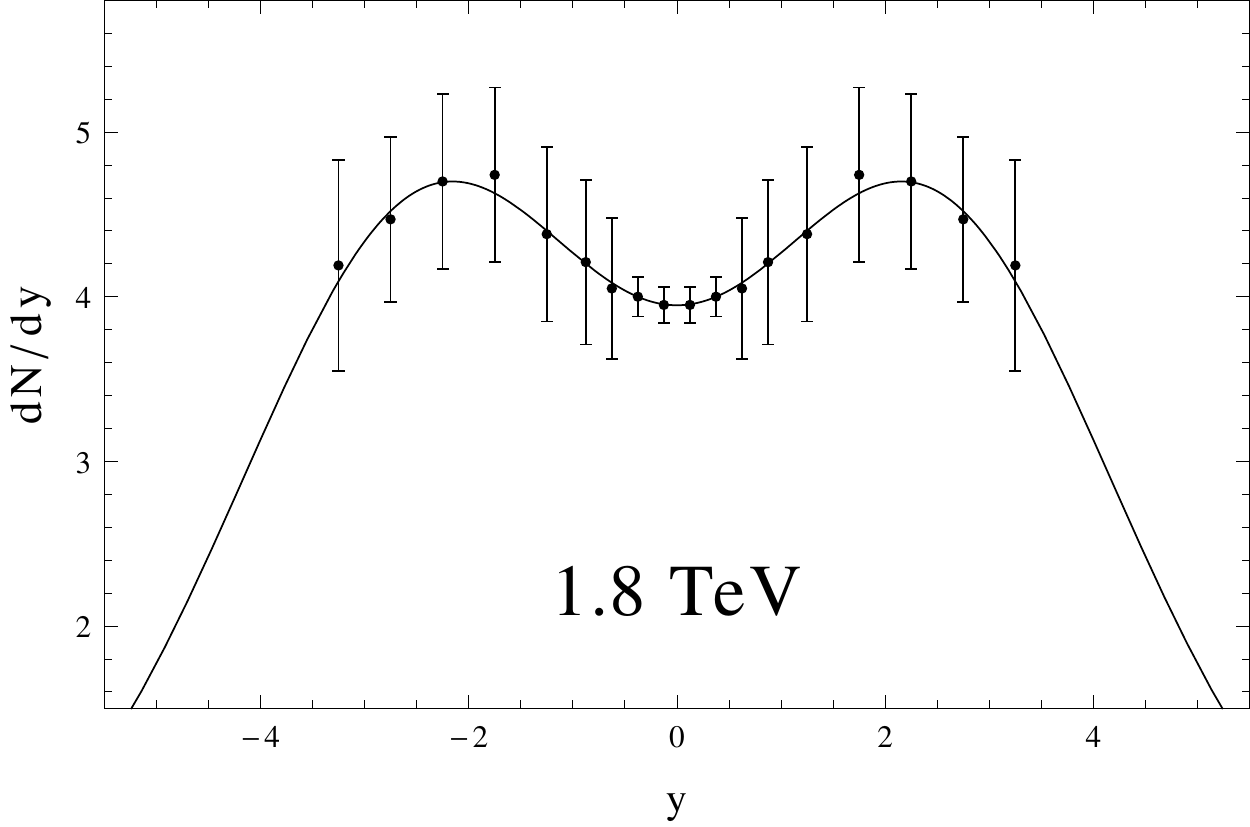}
}
\subfigure[]{
\includegraphics[scale=0.6]{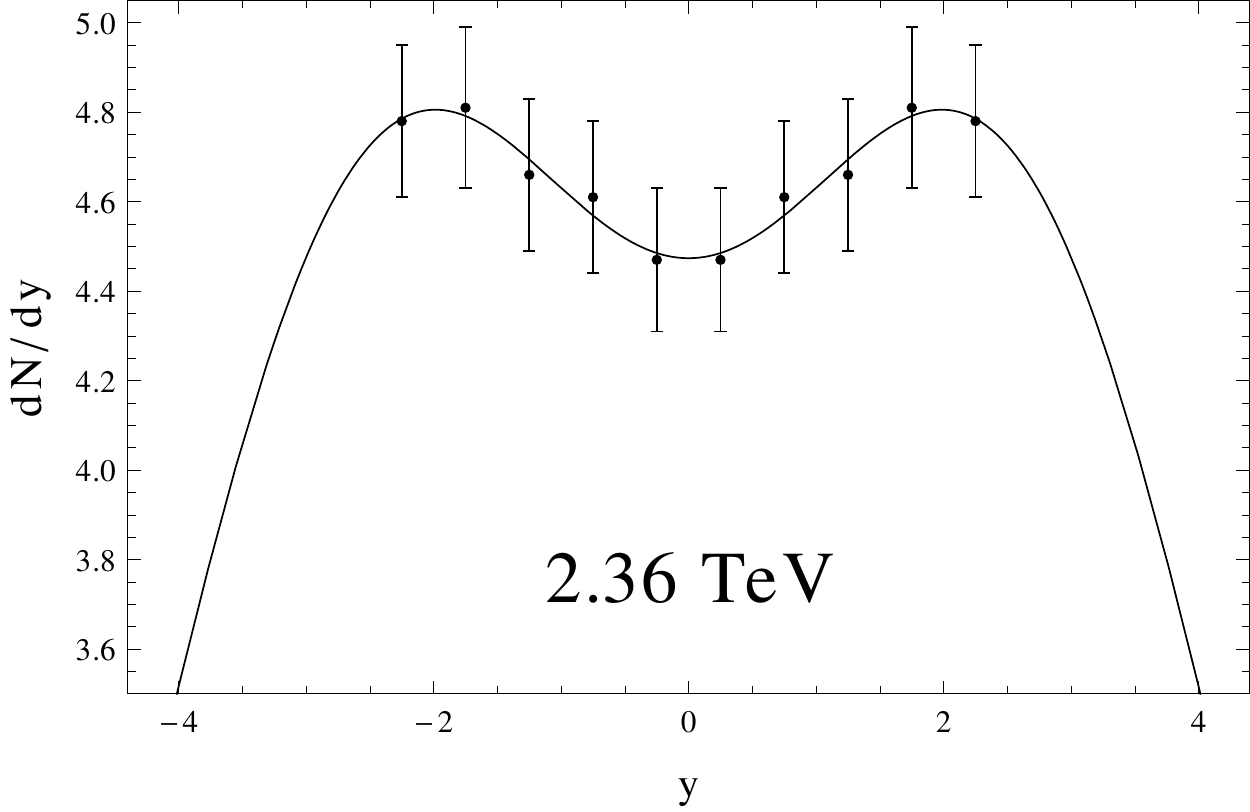}
}
\subfigure[]{
\includegraphics[scale=0.6]{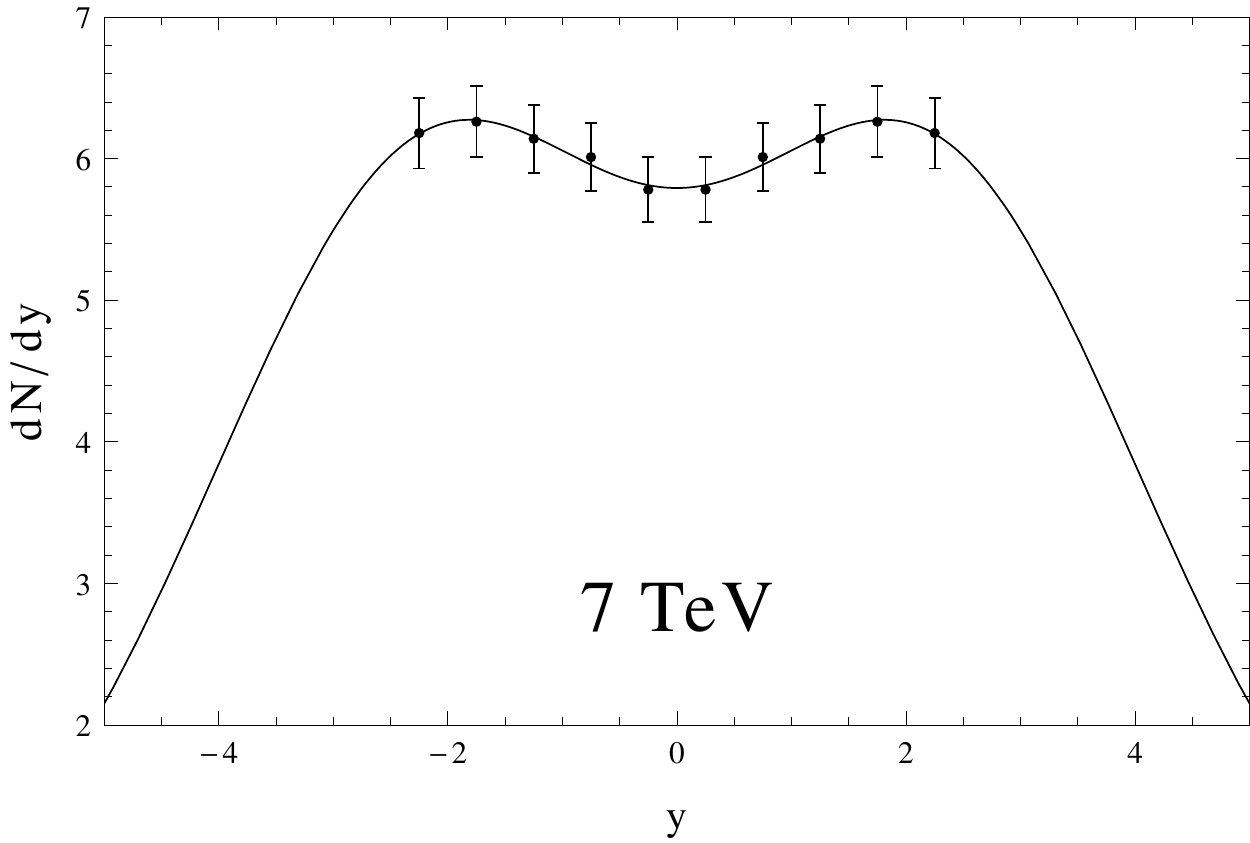}
}
\caption{Series of fittings of experimental data on rapidity distribution (see Table 2) with Eq. \ref{eq:fitformula}. The experimental data are from: (a) UA5 collaboration~\cite{rapidity53GeV} (b) CMS collaboration~\cite{rapidity900e2360GeV}  (c) CDF collaboration~\cite{rapidity1800GeV} (d) CMS collaboration~\cite{rapidity900e2360GeV} (e) CMS collaboration~\cite{rapidity7TeV}}
\label{fig:rapfittings2}
\end{figure}

\cleardoublepage
\newpage
 
 \bibliographystyle{aipproc}

\end{document}